\documentclass[a4paper,fleqn,longmktitlefalse]{cas-dc}

\usepackage[authoryear,sort&compress]{natbib}
\setcitestyle{authoryear,round} 
\setcitestyle{aysep={,},yysep={;}} 

\usepackage{amsmath,amsfonts,bm}

\def\eqref#1{equation~\ref{#1}}

\def\1{\bm{1}}

\DeclareMathAlphabet{\mathsfit}{\encodingdefault}{\sfdefault}{m}{sl}
\SetMathAlphabet{\mathsfit}{bold}{\encodingdefault}{\sfdefault}{bx}{n}

\usepackage[utf8]{inputenc} 
\usepackage[T1]{fontenc}    
\usepackage{hyperref}       
\usepackage{url}            
\usepackage{booktabs}       
\usepackage{amsfonts}       
\usepackage{nicefrac}       
\usepackage{microtype}      
\usepackage{xcolor}         
\usepackage{graphicx}      
\usepackage{subcaption}    
\usepackage{todonotes}
\usepackage{adjustbox}

\usepackage{enumitem}
\usepackage{makecell}

\usepackage{tabularx} 
\usepackage{array}    

\usepackage{lastpage}

\newcommand{\DC}{$DC_{50}$ }
\newcommand{\Dmax}{$D_{max}$ }
\newcommand{\pDC}{$pDC_{50}$ }

\newcommand{\ie}{\textit{i.e.}}
\newcommand{\eg}{\textit{e.g.}}

\def\tsc#1{\csdef{#1}{\textsc{\lowercase{#1}}\xspace}}
\tsc{WGM}
\tsc{QE}

\begin{document}
\let\WriteBookmarks\relax
\def\floatpagepagefraction{1}
\def\textpagefraction{.001}

\shorttitle{Modeling PROTAC Degradation Activity with Machine Learning} 

\shortauthors{Ribes \textit{et al.}}  

\title[mode=title]{Modeling PROTAC Degradation Activity with Machine Learning}  

\affiliation[1]{
    organization={Department of Computer Science and Engineering, Section for Data Science and AI, Chalmers University of Technology},
    addressline={Chalmersplatsen 4}, 
    city={Gothenburg},
    postcode={412 96}, 
    country={Sweden}
}

\affiliation[2]{
    organization={Medicinal Chemistry, Research and Early Development, Respiratory and Immunology (R\&I), ~BioPharmaceuticals R\&D, AstraZeneca},
    addressline={Pepparedsleden 1}, 
    city={Mölndal},
    postcode={431 83}, 
    country={Sweden}
}

\author[1]{Stefano Ribes}[type=author,
      style=european,
      auid=1,
      bioid=1,
      orcid=0009-0009-2774-8792,
]

\ead{ribes@chalmers.se}

\author[2]{Eva Nittinger}[type=editor,
      style=european,
      auid=2,
      bioid=2,
      orcid=0000-0001-7231-7996,
]

\ead{eva.nittinger@astrazeneca.com}

\author[2]{Christian Tyrchan}[type=editor,
      style=european,
      auid=3,
      bioid=3,
      orcid=0000-0002-6470-984X,
]

\ead{christian.tyrchan@astrazeneca.com}

\author[1]{Rocío Mercado}[type=author,
      style=european,
      auid=4,
      bioid=4,
      orcid=0000-0002-6170-6088,
]

\ead{rocio.mercado@chalmers.se}

\cormark[1] 

\cortext[1]{Corresponding author}

\begin{abstract}
PROTACs are a promising therapeutic modality that harnesses the cell's built-in degradation machinery to degrade specific proteins. Despite their potential, developing new PROTACs is challenging and requires significant domain expertise, time, and cost. Meanwhile, machine learning has transformed drug design and development. In this work, we present a strategy for curating open-source PROTAC data and an open-source deep learning tool for predicting the degradation activity of novel PROTAC molecules. The curated dataset incorporates important information such as \pDC, \Dmax, E3 ligase type, POI amino acid sequence, and experimental cell type.
Our model architecture leverages learned embeddings from pretrained machine learning models, in particular for encoding protein sequences and cell type information.
We assessed the quality of the curated data and the generalization ability of our model architecture against new PROTACs and targets via three tailored studies, which we recommend other researchers to use in evaluating their degradation activity models.
In each study, three models predict protein degradation in a majority vote setting, reaching a top test accuracy of 80.8\% and 0.865 ROC AUC, and a test accuracy of 62.3\% and 0.604 ROC AUC when generalizing to novel protein targets.
Our results are not only comparable to state-of-the-art models for protein degradation prediction, but also part of an open-source implementation which is easily reproducible and less computationally complex than existing approaches.
\end{abstract}


\maketitle

\section{Introduction}
\label{sec:introduction}

Machine learning (ML) has transformed various scientific domains, including drug design and discovery, by offering novel solutions to complex, multi-objective optimization challenges \citep{winter2019efficient, atance2022novo, gao2022sample, fromer2023computer}. In the context of medicinal chemistry, ML techniques have revolutionized the process of identifying and optimizing potential drug candidates. Traditionally, drug discovery has relied heavily on trial-and-error experimentation, which is not only time-consuming but also expensive. ML techniques have the potential to significantly accelerate and improve this process by predicting properties of molecules \textit{in silico}, such as binding affinity, solubility, and toxicity, with remarkable accuracy \citep{gorantla2024benchmarking, vassileiou2023unified, born2023chemical}.
This in turn saves time and money in early-stage drug discovery by focusing resources on the most promising candidates.
At the same time, AI models' high performance can potentially lead to better designed drugs for patients.

In order to develop ML models for chemistry, ML algorithms leverage vast datasets containing molecular structures, biological activities, and chemical properties to learn intricate patterns and relationships, also called quantitative structure-activity relationships (QSAR). These algorithms can discern subtle correlations and structure in molecular data that are difficult for human experts to identify. Consequently, ML-based approaches aid in predicting which molecules are likely to be effective drug candidates, thereby narrowing down the search space and saving resources \citep{wu2021we, blaschke2020reinvent}.

\begin{figure*}[t!]
    \centering
    \begin{subfigure}[t]{0.5\textwidth}
        \centering
        \adjustbox{valign=m}{
            \includegraphics[width=0.9\columnwidth]{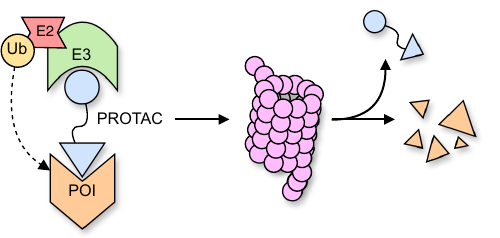}
        }
        \vspace{1em} 
        \caption{}
        \label{fig:protac_mechanism}
    \end{subfigure}%
    \begin{subfigure}[t]{0.5\textwidth}
        \centering
        \adjustbox{valign=m}{
            \includegraphics[width=0.99\columnwidth]{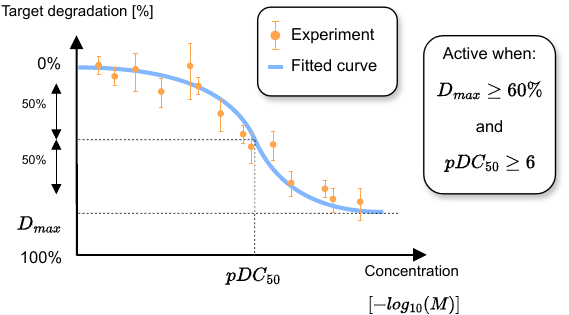}
        }
        \vspace{-0.55em} 
        \caption{}
        \label{fig:dose_response_curve}
    \end{subfigure}%
    \caption{(a) Schematic representation of the PROTAC mechanism of action: the proteasome (violet) degrades the ubitiquinated POI targeted by the PROTAC. After degradation, the PROTAC becomes available again for new targets. (b) Example of a typical PROTAC dose-response curve, along with the activity thresholds used in this work.}
    \label{fig:protac_mechanism_dose_response_curve}
\end{figure*}

PROTACs, or PROteolysis TArgeting Chimeras, represent an innovative class of therapeutic agents with immense potential in challenging disease areas \citep{liu2020protacs, tomoshige2021protacs, hu2022recent}. Unlike traditional small molecule inhibitors, PROTACs operate by harnessing the cell's natural protein degradation machinery, the proteaosome, to eliminate a protein of interest (POI), as summarized in Figure \ref{fig:protac_mechanism}.
This catalytic mechanism of action for targeted protein degradation (TPD) offers several advantages over conventional approaches, which frequently work by having a small molecule drug bind tightly to and thus block a protein's active site.
In fact, by leveraging their unique mechanism, PROTACs bypass the need for tight binding to specific protein pockets, offering a novel strategy for targeting previously ``undruggable'' proteins.
This approach is particularly relevant in cases where inhibiting the target's activity might not be sufficient; notable examples include certain neurodegenerative diseases like Alzheimer's, where misfolded proteins agglomerate and lead to negative downstream effects in patients~\citep{bekes2022protac}.

By catalytically degrading POIs, PROTACs have the potential to offer more comprehensive therapeutic effects at lower doses relative to traditional inhibitors.
Their capacity for TPD highlights the necessity of thorough efficacy evaluations, typically conducted through dose-response assessments (Figure \ref{fig:dose_response_curve}) to determine critical parameters such as \DC (the molar concentration of PROTAC at half maximum degradation of the POI; the lower the better) and \Dmax (the highest percentage of degraded POI; the higher the better) \citep{gesztelyi_hill_2012}.
However, PROTAC development and evaluation face significant challenges due to the limited availability of open-source tools and resources specifically designed for this molecule class, a gap predominantly filled by tools aimed at small molecule inhibitors \citep{nori2022novo, mostofian_targeted_2023}.

To address these challenges, our work introduces a comprehensive machine learning toolkit and curated data specifically designed for PROTAC research. We have developed predictive models that leverage the curated data to effectively forecast the degradation activity of PROTACs, achieving high predictive accuracy and ROC-AUC scores on the test set (top 82.6\% and 0.848, respectively).
Our system, fully open-source and easily accessible via a Python package, is designed to streamline the predictive modeling of PROTAC degradation activity, thus facilitating the rapid evaluation and optimization of new PROTAC designs.
Our contribution significantly expands the available public resources for PROTAC development, setting a new baseline in the application of ML techniques to this emerging therapeutic area.

\section{Materials and Methods}

\subsection{Data Curation}
\label{sec:datacuration}

For this work, we collected and curated data from PROTAC-DB \citep{weng2021protac} and PROTAC-Pedia \citep{protacpedia} that represent, to our knowledge, the two largest open datasets for PROTAC data.
PROTAC-DB contains experimental data, scraped from the scientific literature, for 5,388 PROTACs (as of May 2024; version 2.0).
While the PROTAC-DB allows users to query, filter, and analyze PROTAC data via its online platform (\textit{e.g.}, comparing different compounds based on their \DC and $D_{max}$), its data is not specifically structured for ML models, but rather for online access through its web page. Wrangling the data for use in data-driven models requires significant cleaning and curation.
On the other hand, PROTAC-Pedia provides 1,190 crowd-sourced entries (as of May 2024), with details on PROTACs and their degradation activity.

To prepare the data for our models, we extracted and standardized the following features from the PROTAC-DB and PROTAC-Pedia datasets, where a specific combination of the features corresponds to one experiment: the PROTAC compound, cell line identifier, E3 ligase, POI, and degradation metrics (\DC and $D_{max}$).

Each dataset entry includes the SMILES representation of the PROTAC, which was canonicalized using RDKit \citep{rdkit}.
In PROTAC-DB, cell line information was predominantly found in textual assay descriptions, such as \textit{``degradation in LNCaP cells after 6 h at 0.1/1000/10000 nM''}, with \textit{``LNCaP''} being the cell type in this statement.
Cell type information was extracted using regex parsing, with a few manually cleaned entries.
Afterward, cell line names were standardized using Cellosaurus \citep{bairoch2018cellosaurus} to remove synonyms.
The Uniprot IDs~\citep{uniprot} of E3 ligases and POIs lacking that information were manually web searched and added as text to each entry.

For PROTAC-DB, some of the \DC and \Dmax values were obtained by splitting entries containing information for the same PROTAC on multiple assays.
Duplicates consisting of the same SMILES, POI, E3 ligase, and cell line, but different \DC or \Dmax, were handled by merging them into one entry; we assigned as \DC and \Dmax the geometric mean of their reported \DC and \Dmax values, respectively.
A data sample is labeled as \textit{active} when both its \pDC (\ie, the \DC value expressed in negative $log_{10}$ units) and \Dmax are above their respective predefined threshold values; here we used 6 and $60\%$, respectively.
Effectively, each data point is assigned a binary label indicating degradation activity.

\begin{figure*}[t!]
    \centering
    \includegraphics[width=0.75\textwidth]{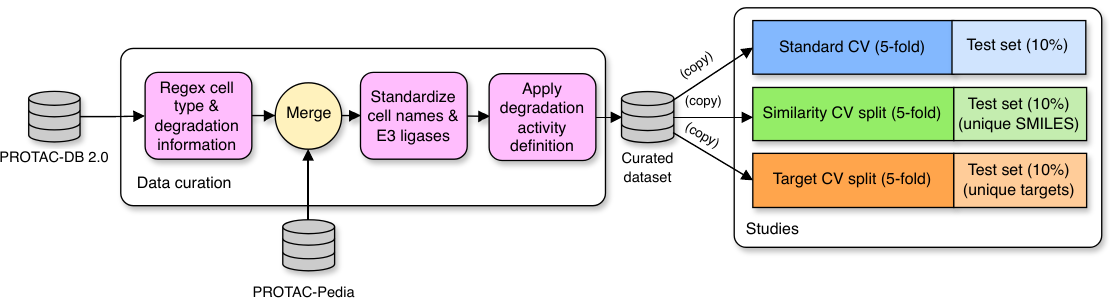}
    \caption{Data curation pipeline and proposed studies.}
    \label{fig:data_curation_pipeline}
\end{figure*}

\subsection{Data Representation}
\label{sec:data_representation}

Given the available data consisting of PROTACs, E3 ligases, POIs, and cell lines, our goal is to encode the diverse information into efficient numerical embeddings that an ML model can leverage.
Because our pool of curated data has a limited size ($\sim 10^3$ data samples), we decided to focus on learning individual embeddings for each of the following: the PROTAC, E3 ligase, POI, and cell type for each experiment. 

For PROTACs, their SMILES strings are converted, via RDKit \citep{rdkit}, to Morgan fingerprints of 256 bits with a radius 10 and stereochemistry information included, with 256 being the smallest $2^n$ vector length not resulting in the overlap of any two fingerprints. We experimented with several bit lengths and radii on the available data while counting the number of overlapping fingerprints, \ie, different PROTACs having identical fingerprints. Ultimately, we selected the combination with the smallest bit length and radius not resulting in any overlapping fingerprints.
The two proteins corresponding to the E3 ligase and POI are converted into pre-computed Uniprot embeddings of 1024 elements \citep{uniprot, bio-embeddings}.
Appendix \ref{appendix:protein_embeddings_as_aa_counts} includes an evaluation using amino acid sequence counts as protein embeddings, for comparison.
Cell lines were one-hot encoded, although other embeddings were also explored and are described in Appendix \ref{appendix:cell_line_embeddings}.
Finally, a pretrained sentence Transformer model \citep{reimers-2019-sentence-bert} was used to encode the text descriptions into numerical embedding vectors of 768 elements.

Once we collected all the embeddings representations, POI and E3 ligase embeddings were normalized independently by removing the respective mean and by scaling to unit variance.
The normalization parameters are learned on the given training set and kept fixed for validation and testing.
Morgan fingerprints and cell line one-hot encodings, being binary vectors, were not normalized.

\subsection{Model Architecture}

\begin{figure*}[t!]
    \centering
    \includegraphics[width=0.8\textwidth]{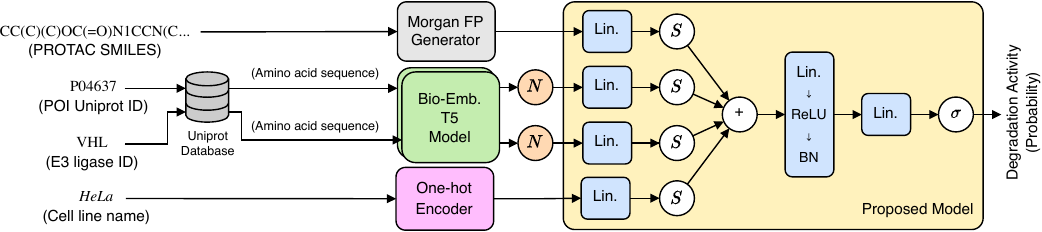}
    \caption{Model pipeline and architecture. The normalization, softmax, and sigmoid functions are denoted as $N$, $S$, and $\sigma$, respectively. The pretrained bio-embedding model can be found in \cite{bio-embeddings}, while the pretrained sentence Transformer is from \cite{reimers-2019-sentence-bert}.}
    \label{fig:model}
\end{figure*}

An illustration of the model architecture is shown in Figure \ref{fig:model}. 
The model includes a set of linear layers, each processing a separate input vector, \ie, the Morgan fingerprints, and the normalized POI, E3 ligase, and cell embeddings, respectively.
The linear layer outputs are then softmax-ed in order to make them of comparable magnitude, and finally summed together.
Lastly, they are forwarded to two additional linear layers, interleaved by a ReLU activation function and a batch norm layer.
The model is trained to optimize a binary cross-entropy loss (with logits).
We set the batch size to 128 and reduce the learning rate by a factor of $10\times$ whenever the validation loss increases compared to the previous training step.
Finally, we apply a sigmoid function to the output of the final linear layer before returning predictions about PROTAC activity.

\subsection{Evaluation Strategy}
\label{sec:evaluation_strategy}

To fully assess the quality of the curated data and the potential performance of DL models in predicting degradation activity, we designed a set of three studies (Figure \ref{fig:data_curation_pipeline}).
In the first study, we seek to identify the potential upper bound of the model performance given the curated data, referred to in this work as the \textit{standard CV split}.
To do so, we randomly pick 10\% of the data as a test set, and leave the remaining data for training with 5-fold cross validation (CV). This leads to an ensemble of five trained models, one per CV fold.
In the next study, we explore model generalization against unseen POIs, referred to in this work as the \textit{target CV split}.
Similar to the previous study, we carefully select 10\% of the available data for testing, such that the POI does not appear in the remaining 90\% of the data which is used for training (5-fold CV). The target CV split hold-out set was selected based on the distribution of active and inactive entries within Uniprot groups, where we prioritized less common POIs first and ensured that adding each group to the test set did not exceed the specified test split proportion while maintaining a balance between active and inactive entries. This is similar to the constraints in the similarity CV split.
Finally, we evaluate the model generalization performance to new PROTACs, referred to in this work as the \textit{similarity CV split}.
To do so, we compute the average Tanimoto distance from all PROTAC Morgan fingerprints to all other PROTAC fingerprints in the full data.
For generating the test set for this experiment, we isolated the data entries starting from the ones where their PROTAC is mapped to a high average Tanimoto distance, until reaching 10\% of the total available data, leaving the rest for CV training.

For each study, we used stratified group CV as implemented in scikit-learn to ensure each fold has a balanced distribution of active and inactive compounds. The validation set performance is obtained by averaging the validation performance on each of the five CV folds using the best set of hyperparameters found via the Optuna optimization framework \citep{akiba2019optuna}. The test performance is obtained from the held-out test set by averaging the performance of three models with the best hyperparameters, each trained using a different random seed on all the data used in CV.

\subsection{Hyperparameter Tuning and Ablation Studies}
\label{sec:hyperparameter_tuning_and_ablation}

\begin{table}[h!]
    \centering
    \caption{Parameters optimized by Optuna: the table reports the parameter name, its type, \ie, categorical (Cat) or continuous (Cont), and the range of values or options suggested in each trial. We apply SMOTE oversampling \citep{chawla2002smote} to the concatenated input data, when suggested.}
    \label{tab:optuna_params}
    \begin{tabular*}{\tblwidth}{@{}lll@{}}
    \toprule
    \textbf{Parameter}            & \textbf{Type}                & \textbf{Options / Range}                            \\ \midrule
    Hidden Dimension     & Cat         & [32, 64, 128, 256, 512]                            \\
    Learning Rate        & Cont (log)  & [$1e^{-5}$, $1e^{-3}$]                     \\
    Use SMOTE            & Cat         & [True, False]                              \\
    SMOTE $k$ Neighbors  & Cat         & [3, 4, ..., 15]                            \\
    \bottomrule
    \end{tabular*}
\end{table}

For hyperparameter tuning we leveraged the Optuna optimization framework \citep{akiba2019optuna}.
In each study, we let Optuna spawn 150 trials to suggest a model architecture and hyperparameters to be used to train the models in the CV folds (we used 5 folds).
Each trial is instructed to sample all the hyperparameters values listed in Table \ref{tab:optuna_params}.
Using Optuna, the goal is to find the best set of hyperparameters that maximize the average validation ROC-AUC score across the CV folds.
The best hyperparameter configuration is then used to train three separate models per study, each with randomly initialized weights (with different seeds), in order to account for model variability.
The best configuration models in each study are trained on the combined study's train and validation sets and evaluated on the respective held-out test set.

Additionally, we conducted an ablation study in which we progressively set input vectors to all zeros, and feed them to the three best models trained during the random split study.

\section{Results and Discussion}

\subsection{Degradation Activity Thresholds}

A data sample was labeled \textit{active} if its \pDC is $\geq6.0$ (equivalent to 1 $\mu M$) and \Dmax  $\geq60\%$.The \pDC threshold helps identify PROTACs with therapeutic potential, as molecules above this threshold are likely to show significant biological activity. Similarly, the \Dmax threshold helps identify PROTACs capable of achieving substantial degradation of the target protein, indicative of efficacy.
\pDC is particularly relevant for drug design, as it allows for the prioritization of compounds that not only bind to the POI but also lead to its effective degradation at a reasonable concentration.
By choosing the above thresholds, we aimed to mitigate model bias, ensuring our dataset includes a balanced representation of both active and inactive compounds, enhancing the model's generalizability.
Note that a PROTAC can be labeled active in one cell type and inactive in another, such as DT2216, a Bcl-xL degrader, which is active in MOLT-4 cancer cells ($pDC_{50}/D_{max} = 7.20/90.8\%$) and inactive in 2T60 hybrid cells ($pDC_{50}/D_{max} = 5.52/26.0\%$) \citep{khan2019selective}.

\subsection{Curated Dataset}

\begin{figure*}[t!]
    \centering
    \begin{subfigure}{0.49\textwidth}
        \centering
        \includegraphics[width=0.9\columnwidth]{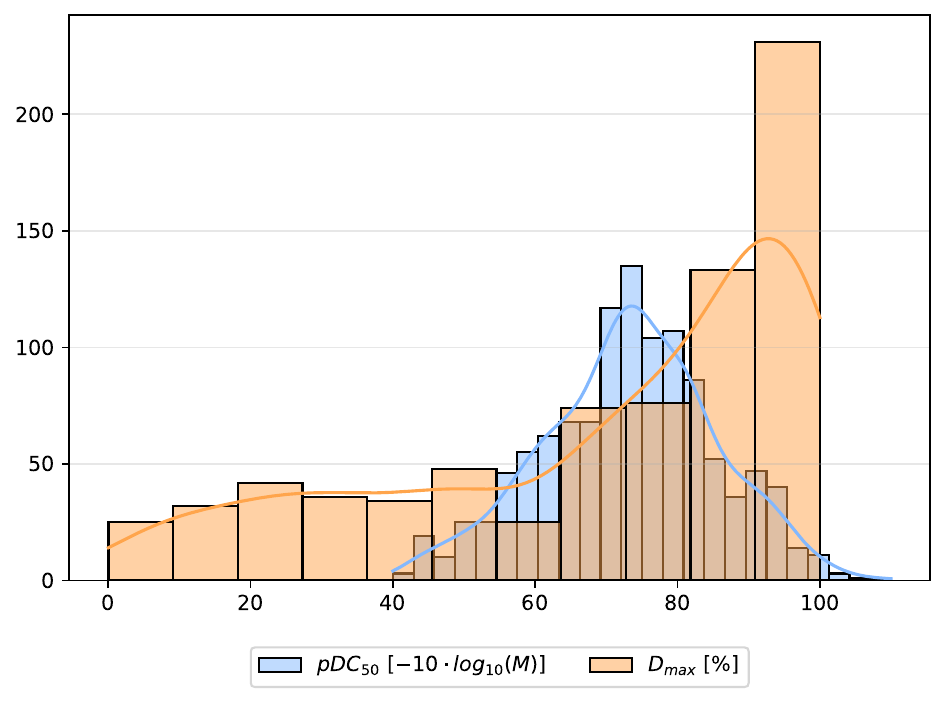}
        \caption{}
        \label{fig:dc50_dmax_distributions}
    \end{subfigure}%
    \begin{subfigure}{0.49\textwidth}
        \centering
        \includegraphics[width=0.9\columnwidth]{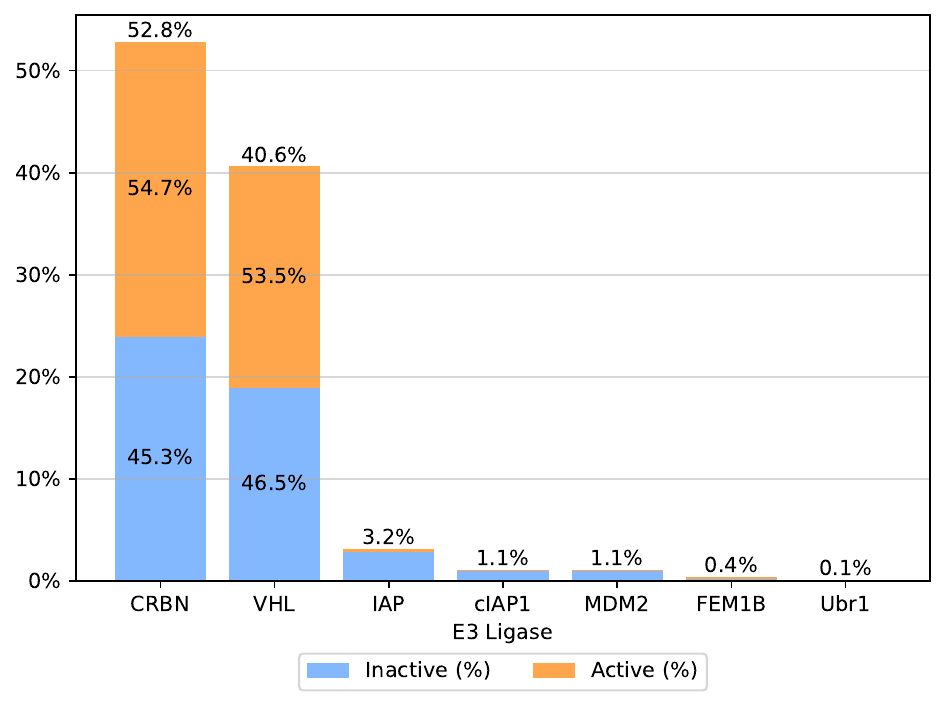}
        \caption{}
        \label{fig:active_inactive_per_e3_ligase}
    \end{subfigure}%
    \caption{(a) Histogram of \pDC and \Dmax in the full curated dataset. Note that the \pDC values are scaled $10\times$ to better display them along side \Dmax values, although they are not bounded by 0 and 100 as \Dmax is. (b) The percentage of curated data associated with each E3 ligase and the active/inactive percentage of data points per E3 ligase.}
    \label{fig:dataset_distributions}
\end{figure*}

After data curation, we were able to extract a total of 2,141 data samples, out of which 812 (37.9\%) report information about $D_{max}$, and 1,350 (63.1\%) include a \DC value.
The curated dataset contained no E3 ligase knockout cell lines, as determined by searching for ``-/-'' in the cell line text (a few other knockouts, however, were identified).
When applying the aforementioned definition of degradation activity, we isolated 759 data samples, 391 (51.52\%) of which are labeled active and the remaining 368 (48.48\%) inactive.
An overview of the distribution of \pDC and \Dmax values is shown in Figure \ref{fig:dc50_dmax_distributions}.
We can see that the majority of the data samples are normally concentrated around the \pDC threshold of 7.31, with a few outliers.
\Dmax values, on the other hand, are more spread out, with half of the samples showing a \Dmax above 80\% (median value).

Figure \ref{fig:active_inactive_per_e3_ligase} shows the distribution of E3 ligases and their frequency in the dataset, together with the percentage of active/inactive samples associated with each of them.
PROTACs are equally distributed (roughly) among the two main E3 ligases, cereblon (CRBN) and von Hippel–Lindau (VHL), with a small fraction of PROTACs being evaluated with other E3 ligases.
We see that CRBN and VHL are indeed the most common (52.8\% and 40.6\%, respectively), whereas 5.6\% of the data samples report less common E3 ligases: IAP (3.16\%), MDM2 (1.10\%), cIAP1 (1.10\%), FEM1B (0.37\%), Ubr1 (0.11\%).
Regarding the active samples distribution among E3 ligases, CRBN and VHL are quite balanced (54.7\% and 53.5\%, respectively), and FEM1B and Ubr1 are mostly associated with active samples. The less common MDM2, IAP, and cIAP1 are mostly associated with inactive samples.

\subsection{Model Performance}

\begin{figure*}[t!]
    \centering
    \begin{subfigure}{0.65\textwidth}
        \centering
        \adjustbox{valign=m}{
            \includegraphics[width=0.99\columnwidth]{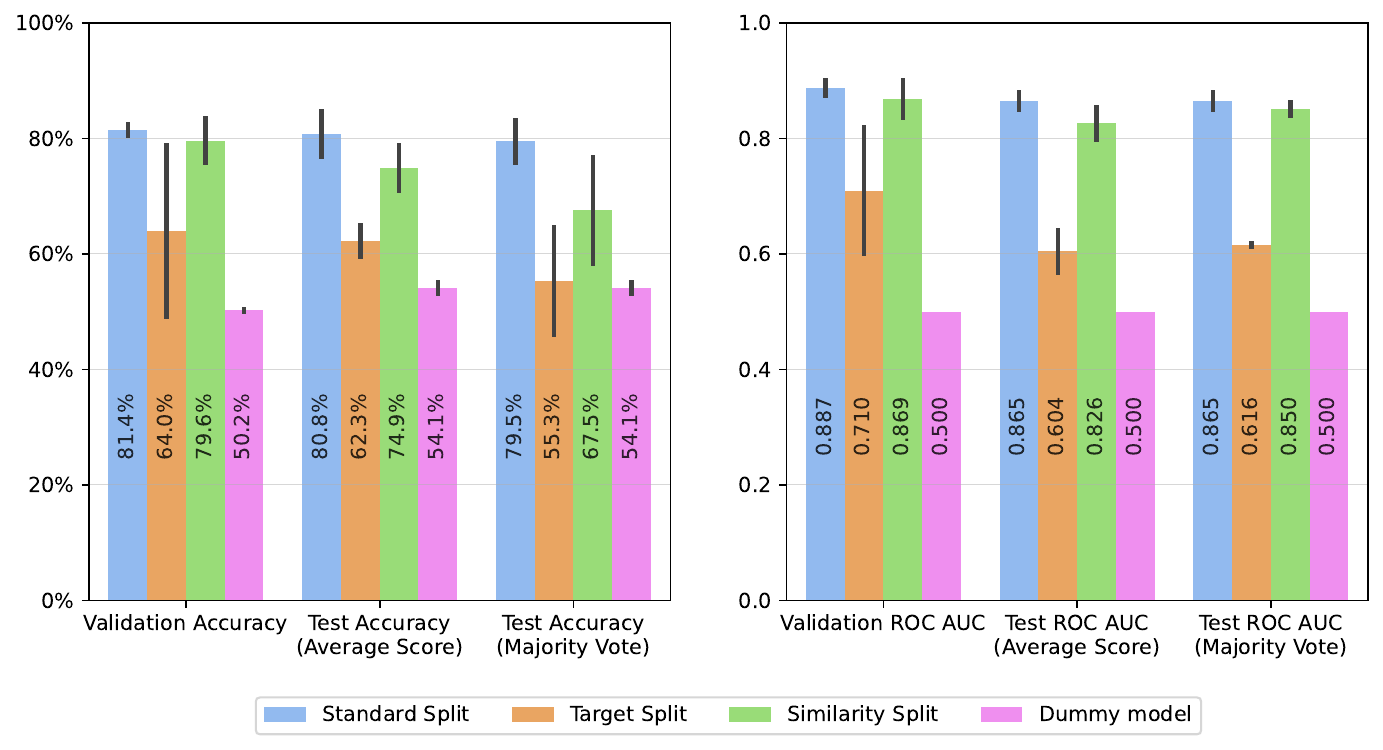}
        }
        \caption{}
        \label{fig:performance}
    \end{subfigure}%
    \begin{subfigure}{0.35\textwidth}
        \centering
        \adjustbox{valign=m}{
            \includegraphics[width=0.999\columnwidth]{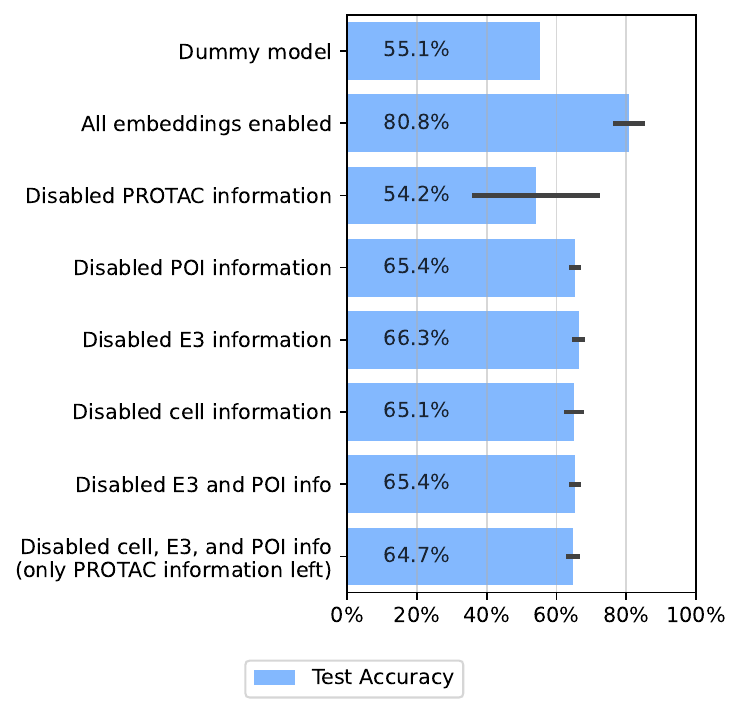}
        }
        \vspace{.5em}
        \caption{}
        \label{fig:ablation}
    \end{subfigure}%
    \caption{(a) Performance of the different models across the various studies, with model accuracy plotted on the left and ROC AUC plotted on the right. (b) Ablation results for the standard cross-validation split. Each bar indicates the embedding(s) not available to the model to process.}
    \label{fig:ablation_performance}
\end{figure*}

Figure \ref{fig:performance} reports the performance of the different models across the various studies.
For each study, named after either the \textit{standard}, \textit{target}, or \textit{similarity} split used, we show the mean validation accuracy and ROC-AUC scores of the five models trained during CV (one model per fold) with the best hyperparameters found.
Additionally, the plots show the performance on the test set of three models trained per study with the best hyperparameters found in CV and different initial weights.
For those models, we also report the mean of the test accuracy and ROC-AUC scores, alongside  the test accuracy and ROC-AUC scores calculated using majority voting.
A dummy model is included as a baseline, which always predicts the majority class in the training set.

The performance metrics derived from the standard CV split offer an upper bound for our model's capability, with a validation average/test average/test majority vote accuracy of 81.4\%/80.8\%/79.5\% and a validation average/test average/test majority vote ROC AUC of 0.887/0.865/0.865. These results suggest an optimal scenario where the model has access to a diverse and representative sample of the data during training, maximizing its learning potential. The standard split serves as an upper bound estimate for model performance, as real-life scenarios generally require more constrained and specialized testing conditions.

On the other hand, in the similarity CV split study, designed to evaluate the model's generalizability to unseen PROTAC compounds that do not share structural similarities with the training set, our model reached a remarkable validation average/test average/test majority vote accuracy of 79.6\%/74.9\%/67.5\% and a validation average/test average/test majority vote ROC AUC of 0.869/0.826/0.850. The high performance in this study indicates the model's robust ability to extrapolate from known PROTACs to predict the activities of novel molecules.

Finally, the target split study presents a significant challenge for our model, as evidenced by the lower validation average/test average/test majority vote accuracy of 64.0\%/62.3\%/55.3\% and a validation average/test average/test majority vote ROC AUC of 0.710/0.604/0.616. This study tests the model's ability to generalize across different protein targets, a critical factor for PROTAC design in novel disease mechanisms. The diminished performance suggests a need for improved protein representations or for embeddings that better capture more detailed and relevant features of the target proteins. Moreover, it underscores the necessity for more extensive and diverse datasets that include a broader array of PROTACs and targets.

Additional performance metrics are reported in Appendix \ref{appendix:scores}.
Appendix \ref{appendix:xgboost} includes instead the performance scores of an XGBoost model evaluated on the aforementioned studies \citep{Chen_2016}.

\subsection{Ablation Studies}

The ablation study summarized in Figure \ref{fig:ablation} highlights the contributions of various embeddings to model performance in PROTAC activity prediction.
We focus on the average test accuracy of the three models trained with the best hyperparameters in the standard split study. With all embeddings enabled, the three models achieved an average test accuracy of 80.8\%, serving as the baseline for full-feature utilization. Disabling cell, E3 ligase, and protein of interest (POI) embeddings individually led to varied decreases in performance, with test accuracies of 65.1\%, 66.3\%, and 65.4\%, respectively. This highlights the importance of each type of embedding in enhancing predictive accuracy.

Notably, the model performance dropped below that of the dummy model when disabling compound information, emphasizing the importance of the PROTAC fingerprints.
This is further highlighted by the test accuracy of the combination of disabled POI, E3, and cell embeddings (leaving the PROTAC information only), which reached 64.7\%, close to other setups in which only a single component was disabled.
In general, molecular fingerprints appear to be the most relevant input feature to the model.
However, the general trend of high accuracy drops suggests that the contextual embeddings collectively contribute with significant predictive value beyond the structural information provided by molecular fingerprints alone.

Overall, this ablation study demonstrates the synergistic effect of integrating diverse embeddings, including compound structure (PROTAC fingerprint) and biological context (cell type, E3 ligase, POI), to capture the diverse determinants of biological activity in PROTACs.

\section{Related Work}

The studies most closely aligned with our work are those of \cite{li_deepprotacs_2022} and \cite{nori2022novo}. \cite{li_deepprotacs_2022} introduces DeepPROTACs, a deep learning model for prognosticating PROTAC activity, whereas \cite{nori2022novo} proposes instead a LightGBM model for predicting protein degradation activity.
LightGBM is a gradient boosting framework that uses a histogram-based approach for efficient, high-performance ML tasks~\citep{ke2017lightgbm}.

The DeepPROTACs architecture encompasses multiple branches employing long short-term memory (LSTM) and graph neural network (GNN) components, all combined prior to a prediction head. Each branch processes distinct facets of the ternary complex, encompassing elements like E3 ligase and POI binding pockets, along with the individual components of the PROTAC: the warhead, linker, and E3 ligand. The model's performance culminates in an average prediction accuracy of 77.95\% and a ROC-AUC score of 0.8470 on a validation set drawn from the PROTAC-DB. The LightGBM model, on the other hand, achieves a ROC-AUC of 0.877 on a PROTAC-DB test set with a much simpler model architecture and input representation.

Notwithstanding their achievements, the DeepPROTACs and LightGBM models both exhibit certain limitations. In DeepPROTACs, there is a potential risk of information loss as the PROTAC SMILES are partitioned into their constituent E3 ligands, warheads, and linkers, which are then fed into separate branches of the model. Secondly, while the authors undertake advanced molecular docking of the entire PROTAC-POI-E3 ligase complex, their subsequent focus on the 3D binding pockets of the POI and E3 ligase renders it less amenable for experimental replication and practical use.
Finally, and perhaps most importantly, the potential for data leakage during hyperparameter optimization and its effects on out-of-distribution (OOD) generalization was not investigated.
Data leakage between the different PROTAC components in the training and test sets of the model may artificially render a more accurate model that does not generalize well to new real-word data, necessitating more rigorous testing procedures.
Because of that, generalization of the DeepPROTACs model would need to be further investigated on a separate test set.

\section{Conclusions}

In this work, we curated open-source PROTAC data and introduced a versatile toolkit for predicting PROTAC degradation effectiveness in three different experimental scenarios, aiming to assess the quality of our curated data and model generalizability.
The performance of our models, achieving a top 80.8\% test accuracy and a 0.865 ROC-AUC test score are competitive with, if not surpassing, existing methods for protein degradation prediction. Ours are also the first models to consider both \DC and \Dmax in predicting degradation activity for PROTACs, a significant contribution as both properties are important to determining PROTAC efficacy.
We show that our models can generalize well to unseen PROTACs, while struggling with unseen targets, highlighting the need for more comprehensive protein representations and more extensive datasets.
Finally, our approach offers open-source accessibility, ease of reproducibility, and a less computationally complex alternative to previous work, making it a valuable resource for researchers working on data-driven approaches to PROTAC engineering.

\subsection*{Reproducibility Statement and Code Availability}

Code for this work is available at \url{https://github.com/ribesstefano/PROTAC-Degradation-Predictor}.
The repository contains detailed instructions to reproduce the results presented in this work, including: the dataset curation process and the curated datasets, hyperparameter tuning, and the Optuna studies which can be used to train/retrieve all models presented herein.
Additionally, we developed a web application (with GUI) hosted at: \url{https://huggingface.co/spaces/ailab-bio/PROTAC-Degradation-Predictor}.

\subsection*{Acknowledgments}
The authors acknowledge computing resources from AstraZeneca and the National Academic Infrastructure for Supercomputing in Sweden (NAISS), partially funded by the Swedish Research Council through grant agreement no. 2022-06725. SR thanks the Health Engineering Area of Advance and the Gender Initiative for Excellence (Genie) for funding. RM thanks the Wallenberg AI, Autonomous Systems, and Software Program (WASP) for funding.

\bibliographystyle{cas-model2-names}
\bibliography{references.bib}

\begin{thebibliography}{30}
\expandafter\ifx\csname natexlab\endcsname\relax\def\natexlab#1{#1}\fi
\providecommand{\url}[1]{\texttt{#1}}
\providecommand{\href}[2]{#2}
\providecommand{\path}[1]{#1}
\providecommand{\DOIprefix}{doi:}
\providecommand{\ArXivprefix}{arXiv:}
\providecommand{\URLprefix}{URL: }
\providecommand{\Pubmedprefix}{pmid:}
\providecommand{\doi}[1]{\href{http://dx.doi.org/#1}{\path{#1}}}
\providecommand{\Pubmed}[1]{\href{pmid:#1}{\path{#1}}}
\providecommand{\bibinfo}[2]{#2}
\ifx\xfnm\relax \def\xfnm[#1]{\unskip,\space#1}\fi
\bibitem[{Akiba et~al.(2019)Akiba, Sano, Yanase, Ohta and Koyama}]{akiba2019optuna}
\bibinfo{author}{Akiba, T.}, \bibinfo{author}{Sano, S.}, \bibinfo{author}{Yanase, T.}, \bibinfo{author}{Ohta, T.}, \bibinfo{author}{Koyama, M.}, \bibinfo{year}{2019}.
\newblock \bibinfo{title}{Optuna: A next-generation hyperparameter optimization framework}, in: \bibinfo{booktitle}{Proceedings of the 25th ACM SIGKDD International Conference on Knowledge Discovery \& Data Mining}, pp. \bibinfo{pages}{2623--2631}.
\bibitem[{Atance et~al.(2022)Atance, Diez, Engkvist, Olsson and Mercado}]{atance2022novo}
\bibinfo{author}{Atance, S.R.}, \bibinfo{author}{Diez, J.V.}, \bibinfo{author}{Engkvist, O.}, \bibinfo{author}{Olsson, S.}, \bibinfo{author}{Mercado, R.}, \bibinfo{year}{2022}.
\newblock \bibinfo{title}{De novo drug design using reinforcement learning with graph-based deep generative models}.
\newblock \bibinfo{journal}{Journal of Chemical Information and Modeling} \bibinfo{volume}{62}, \bibinfo{pages}{4863--4872}.
\bibitem[{Bairoch(2018)}]{bairoch2018cellosaurus}
\bibinfo{author}{Bairoch, A.}, \bibinfo{year}{2018}.
\newblock \bibinfo{title}{{The Cellosaurus, a Cell-Line Knowledge Resource}}.
\newblock \bibinfo{journal}{Journal of Biomolecular Techniques} \bibinfo{volume}{29}, \bibinfo{pages}{25}.
\bibitem[{B{\'e}k{\'e}s et~al.(2022)B{\'e}k{\'e}s, Langley and Crews}]{bekes2022protac}
\bibinfo{author}{B{\'e}k{\'e}s, M.}, \bibinfo{author}{Langley, D.R.}, \bibinfo{author}{Crews, C.M.}, \bibinfo{year}{2022}.
\newblock \bibinfo{title}{{PROTAC targeted protein degraders: the past is prologue}}.
\newblock \bibinfo{journal}{Nature Reviews Drug Discovery} \bibinfo{volume}{21}, \bibinfo{pages}{181--200}.
\bibitem[{Blaschke et~al.(2020)Blaschke, Ar{\'u}s-Pous, Chen, Margreitter, Tyrchan, Engkvist, Papadopoulos and Patronov}]{blaschke2020reinvent}
\bibinfo{author}{Blaschke, T.}, \bibinfo{author}{Ar{\'u}s-Pous, J.}, \bibinfo{author}{Chen, H.}, \bibinfo{author}{Margreitter, C.}, \bibinfo{author}{Tyrchan, C.}, \bibinfo{author}{Engkvist, O.}, \bibinfo{author}{Papadopoulos, K.}, \bibinfo{author}{Patronov, A.}, \bibinfo{year}{2020}.
\newblock \bibinfo{title}{{REINVENT 2.0: an AI tool for de novo drug design}}.
\newblock \bibinfo{journal}{Journal of Chemical Information and Modeling} \bibinfo{volume}{60}, \bibinfo{pages}{5918--5922}.
\bibitem[{Born et~al.(2023)Born, Markert, Janakarajan, Kimber, Volkamer, Mart{\'\i}nez and Manica}]{born2023chemical}
\bibinfo{author}{Born, J.}, \bibinfo{author}{Markert, G.}, \bibinfo{author}{Janakarajan, N.}, \bibinfo{author}{Kimber, T.B.}, \bibinfo{author}{Volkamer, A.}, \bibinfo{author}{Mart{\'\i}nez, M.R.}, \bibinfo{author}{Manica, M.}, \bibinfo{year}{2023}.
\newblock \bibinfo{title}{Chemical representation learning for toxicity prediction}.
\newblock \bibinfo{journal}{Digital Discovery} \bibinfo{volume}{2}, \bibinfo{pages}{674--691}.
\bibitem[{Chawla et~al.(2002)Chawla, Bowyer, Hall and Kegelmeyer}]{chawla2002smote}
\bibinfo{author}{Chawla, N.V.}, \bibinfo{author}{Bowyer, K.W.}, \bibinfo{author}{Hall, L.O.}, \bibinfo{author}{Kegelmeyer, W.P.}, \bibinfo{year}{2002}.
\newblock \bibinfo{title}{{SMOTE: Synthetic Minority Over-sampling TEchnique}}.
\newblock \bibinfo{journal}{Journal of Artificial Intelligence Research} \bibinfo{volume}{16}, \bibinfo{pages}{321--357}.
\bibitem[{Chen and Guestrin(2016)}]{Chen_2016}
\bibinfo{author}{Chen, T.}, \bibinfo{author}{Guestrin, C.}, \bibinfo{year}{2016}.
\newblock \bibinfo{title}{{XGBoost: A Scalable Tree Boosting System}}, in: \bibinfo{booktitle}{Proceedings of the 22nd ACM SIGKDD International Conference on Knowledge Discovery and Data Mining}, \bibinfo{publisher}{ACM}.
\newblock \URLprefix \url{http://dx.doi.org/10.1145/2939672.2939785}, \DOIprefix\doi{10.1145/2939672.2939785}.
\bibitem[{Dallago et~al.(2021)Dallago, Schütze, Heinzinger, Olenyi, Littmann, Lu, Yang, Min, Yoon, Morton and Rost}]{bio-embeddings}
\bibinfo{author}{Dallago, C.}, \bibinfo{author}{Schütze, K.}, \bibinfo{author}{Heinzinger, M.}, \bibinfo{author}{Olenyi, T.}, \bibinfo{author}{Littmann, M.}, \bibinfo{author}{Lu, A.X.}, \bibinfo{author}{Yang, K.K.}, \bibinfo{author}{Min, S.}, \bibinfo{author}{Yoon, S.}, \bibinfo{author}{Morton, J.T.}, \bibinfo{author}{Rost, B.}, \bibinfo{year}{2021}.
\newblock \bibinfo{title}{Learned embeddings from deep learning to visualize and predict protein sets}.
\newblock \bibinfo{journal}{Current Protocols} \bibinfo{volume}{1}, \bibinfo{pages}{e113}.
\newblock \URLprefix \url{https://currentprotocols.onlinelibrary.wiley.com/doi/abs/10.1002/cpz1.113}, \DOIprefix\doi{https://doi.org/10.1002/cpz1.113}.
\bibitem[{{EMBL-EBI}(2023)}]{uniprot}
\bibinfo{author}{{EMBL-EBI}}, \bibinfo{year}{2023}.
\newblock \bibinfo{title}{{UniProt}}.
\newblock \URLprefix \url{https://www.uniprot.org/}.
\bibitem[{Fromer and Coley(2023)}]{fromer2023computer}
\bibinfo{author}{Fromer, J.C.}, \bibinfo{author}{Coley, C.W.}, \bibinfo{year}{2023}.
\newblock \bibinfo{title}{Computer-aided multi-objective optimization in small molecule discovery}.
\newblock \bibinfo{journal}{Patterns} \bibinfo{volume}{4}.
\bibitem[{Gao et~al.(2022)Gao, Fu, Sun and Coley}]{gao2022sample}
\bibinfo{author}{Gao, W.}, \bibinfo{author}{Fu, T.}, \bibinfo{author}{Sun, J.}, \bibinfo{author}{Coley, C.}, \bibinfo{year}{2022}.
\newblock \bibinfo{title}{Sample efficiency matters: a benchmark for practical molecular optimization}.
\newblock \bibinfo{journal}{Advances in Neural Information Processing Systems} \bibinfo{volume}{35}, \bibinfo{pages}{21342--21357}.
\bibitem[{Gesztelyi et~al.(2012)Gesztelyi, Zsuga, Kemeny-Beke, Varga, Juhasz and Tosaki}]{gesztelyi_hill_2012}
\bibinfo{author}{Gesztelyi, R.}, \bibinfo{author}{Zsuga, J.}, \bibinfo{author}{Kemeny-Beke, A.}, \bibinfo{author}{Varga, B.}, \bibinfo{author}{Juhasz, B.}, \bibinfo{author}{Tosaki, A.}, \bibinfo{year}{2012}.
\newblock \bibinfo{title}{The {Hill} equation and the origin of quantitative pharmacology}.
\newblock \bibinfo{journal}{Archive for History of Exact Sciences} \bibinfo{volume}{66}, \bibinfo{pages}{427--438}.
\newblock \URLprefix \url{https://doi.org/10.1007/s00407-012-0098-5}, \DOIprefix\doi{10.1007/s00407-012-0098-5}.
\bibitem[{Gorantla et~al.(2024)Gorantla, Kubincova, Suutari, Cossins and Mey}]{gorantla2024benchmarking}
\bibinfo{author}{Gorantla, R.}, \bibinfo{author}{Kubincova, A.}, \bibinfo{author}{Suutari, B.}, \bibinfo{author}{Cossins, B.P.}, \bibinfo{author}{Mey, A.S.}, \bibinfo{year}{2024}.
\newblock \bibinfo{title}{Benchmarking active learning protocols for ligand-binding affinity prediction}.
\newblock \bibinfo{journal}{Journal of Chemical Information and Modeling} \bibinfo{volume}{64}, \bibinfo{pages}{1955--1965}.
\bibitem[{Hu and Crews(2022)}]{hu2022recent}
\bibinfo{author}{Hu, Z.}, \bibinfo{author}{Crews, C.M.}, \bibinfo{year}{2022}.
\newblock \bibinfo{title}{{Recent Developments in PROTAC-Mediated Protein Degradation: From Bench to Clinic}}.
\newblock \bibinfo{journal}{ChemBioChem} \bibinfo{volume}{23}, \bibinfo{pages}{e202100270}.
\bibitem[{Ke et~al.(2017)Ke, Meng, Finley, Wang, Chen, Ma, Ye and Liu}]{ke2017lightgbm}
\bibinfo{author}{Ke, G.}, \bibinfo{author}{Meng, Q.}, \bibinfo{author}{Finley, T.}, \bibinfo{author}{Wang, T.}, \bibinfo{author}{Chen, W.}, \bibinfo{author}{Ma, W.}, \bibinfo{author}{Ye, Q.}, \bibinfo{author}{Liu, T.Y.}, \bibinfo{year}{2017}.
\newblock \bibinfo{title}{{LightGBM}: A highly efficient gradient boosting decision tree}.
\newblock \bibinfo{journal}{Advances in Neural Information Processing Systems} \bibinfo{volume}{30}.
\bibitem[{Khan et~al.(2019)Khan, Zhang, Lv, Zhang, He, Zhang, Liu, Thummuri, Yuan, Wiegand et~al.}]{khan2019selective}
\bibinfo{author}{Khan, S.}, \bibinfo{author}{Zhang, X.}, \bibinfo{author}{Lv, D.}, \bibinfo{author}{Zhang, Q.}, \bibinfo{author}{He, Y.}, \bibinfo{author}{Zhang, P.}, \bibinfo{author}{Liu, X.}, \bibinfo{author}{Thummuri, D.}, \bibinfo{author}{Yuan, Y.}, \bibinfo{author}{Wiegand, J.S.}, et~al., \bibinfo{year}{2019}.
\newblock \bibinfo{title}{{A selective BCL-XL PROTAC degrader achieves safe and potent antitumor activity}}.
\newblock \bibinfo{journal}{Nature Medicine} \bibinfo{volume}{25}, \bibinfo{pages}{1938--1947}.
\bibitem[{Landrum(2010)}]{rdkit}
\bibinfo{author}{Landrum, G.}, \bibinfo{year}{2010}.
\newblock \bibinfo{title}{rdkit.{Chem}.rdmolops module — {The} {RDKit} 2023.03.1 {Documentation}}.
\newblock \URLprefix \url{https://www.rdkit.org/docs/source/rdkit.Chem.rdmolops.html}.
\bibitem[{Li et~al.(2022)Li, Hu, Zhang, Sun, Liu, Wu, Tian, Ma, Dai, Yang, Gao and Bai}]{li_deepprotacs_2022}
\bibinfo{author}{Li, F.}, \bibinfo{author}{Hu, Q.}, \bibinfo{author}{Zhang, X.}, \bibinfo{author}{Sun, R.}, \bibinfo{author}{Liu, Z.}, \bibinfo{author}{Wu, S.}, \bibinfo{author}{Tian, S.}, \bibinfo{author}{Ma, X.}, \bibinfo{author}{Dai, Z.}, \bibinfo{author}{Yang, X.}, \bibinfo{author}{Gao, S.}, \bibinfo{author}{Bai, F.}, \bibinfo{year}{2022}.
\newblock \bibinfo{title}{{DeepPROTACs} is a deep learning-based targeted degradation predictor for {PROTACs}}.
\newblock \bibinfo{journal}{Nature Communications} \bibinfo{volume}{13}, \bibinfo{pages}{7133}.
\newblock \URLprefix \url{https://www.nature.com/articles/s41467-022-34807-3}, \DOIprefix\doi{10.1038/s41467-022-34807-3}.
\bibitem[{Liu et~al.(2020)Liu, Ma, Liu, Xia, Li, Wang and Wei}]{liu2020protacs}
\bibinfo{author}{Liu, J.}, \bibinfo{author}{Ma, J.}, \bibinfo{author}{Liu, Y.}, \bibinfo{author}{Xia, J.}, \bibinfo{author}{Li, Y.}, \bibinfo{author}{Wang, Z.P.}, \bibinfo{author}{Wei, W.}, \bibinfo{year}{2020}.
\newblock \bibinfo{title}{{PROTACs: a novel strategy for cancer therapy}}, in: \bibinfo{booktitle}{Seminars in Cancer Biology}, \bibinfo{organization}{Elsevier}. pp. \bibinfo{pages}{171--179}.
\bibitem[{London and Prilusky()}]{protacpedia}
\bibinfo{author}{London, N.}, \bibinfo{author}{Prilusky, J.}, .
\newblock \bibinfo{title}{{PROTACpedia}}.
\newblock \bibinfo{howpublished}{\url{https://protacpedia.weizmann.ac.il/}}.
\newblock \bibinfo{note}{Accessed: 2024-05-21}.
\bibitem[{McInnes et~al.(2020)McInnes, Healy and Melville}]{mcinnes2020umapuniformmanifoldapproximation}
\bibinfo{author}{McInnes, L.}, \bibinfo{author}{Healy, J.}, \bibinfo{author}{Melville, J.}, \bibinfo{year}{2020}.
\newblock \bibinfo{title}{{UMAP: Uniform Manifold Approximation and Projection for Dimension Reduction}}.
\newblock \URLprefix \url{https://arxiv.org/abs/1802.03426}, \href{http://arxiv.org/abs/1802.03426}{\tt arXiv:1802.03426}.
\bibitem[{Mostofian et~al.(2023)Mostofian, Martin, Razavi, Patel, Allen, Sherman and Izaguirre}]{mostofian_targeted_2023}
\bibinfo{author}{Mostofian, B.}, \bibinfo{author}{Martin, H.J.}, \bibinfo{author}{Razavi, A.}, \bibinfo{author}{Patel, S.}, \bibinfo{author}{Allen, B.}, \bibinfo{author}{Sherman, W.}, \bibinfo{author}{Izaguirre, J.A.}, \bibinfo{year}{2023}.
\newblock \bibinfo{title}{Targeted {Protein} {Degradation}: {Advances}, {Challenges}, and {Prospects} for {Computational} {Methods}}.
\newblock \bibinfo{journal}{Journal of Chemical Information and Modeling} \bibinfo{volume}{63}, \bibinfo{pages}{5408--5432}.
\newblock \URLprefix \url{https://doi.org/10.1021/acs.jcim.3c00603}, \DOIprefix\doi{10.1021/acs.jcim.3c00603}.
\bibitem[{Nori et~al.(2022)Nori, Coley and Mercado}]{nori2022novo}
\bibinfo{author}{Nori, D.}, \bibinfo{author}{Coley, C.W.}, \bibinfo{author}{Mercado, R.}, \bibinfo{year}{2022}.
\newblock \bibinfo{title}{{De novo PROTAC design using graph-based deep generative models}}.
\newblock \bibinfo{journal}{arXiv preprint arXiv:2211.02660} .
\bibitem[{Reimers and Gurevych(2019)}]{reimers-2019-sentence-bert}
\bibinfo{author}{Reimers, N.}, \bibinfo{author}{Gurevych, I.}, \bibinfo{year}{2019}.
\newblock \bibinfo{title}{{Sentence-BERT: Sentence Embeddings using Siamese BERT-Networks}}, in: \bibinfo{booktitle}{Proceedings of the 2019 Conference on Empirical Methods in Natural Language Processing}, \bibinfo{publisher}{Association for Computational Linguistics}.
\newblock \URLprefix \url{https://arxiv.org/abs/1908.10084}.
\bibitem[{Tomoshige and Ishikawa(2021)}]{tomoshige2021protacs}
\bibinfo{author}{Tomoshige, S.}, \bibinfo{author}{Ishikawa, M.}, \bibinfo{year}{2021}.
\newblock \bibinfo{title}{{PROTACs and other chemical protein degradation technologies for the treatment of neurodegenerative disorders}}.
\newblock \bibinfo{journal}{Angewandte Chemie International Edition} \bibinfo{volume}{60}, \bibinfo{pages}{3346--3354}.
\bibitem[{Vassileiou et~al.(2023)Vassileiou, Robertson, Wareham, Soundaranathan, Ottoboni, Florence, Hartwig and Johnston}]{vassileiou2023unified}
\bibinfo{author}{Vassileiou, A.D.}, \bibinfo{author}{Robertson, M.N.}, \bibinfo{author}{Wareham, B.G.}, \bibinfo{author}{Soundaranathan, M.}, \bibinfo{author}{Ottoboni, S.}, \bibinfo{author}{Florence, A.J.}, \bibinfo{author}{Hartwig, T.}, \bibinfo{author}{Johnston, B.F.}, \bibinfo{year}{2023}.
\newblock \bibinfo{title}{A unified ml framework for solubility prediction across organic solvents}.
\newblock \bibinfo{journal}{Digital Discovery} \bibinfo{volume}{2}, \bibinfo{pages}{356--367}.
\bibitem[{Weng et~al.(2021)Weng, Shen, Cao, Gao, Dong, He, Yang, Li, Wu and Hou}]{weng2021protac}
\bibinfo{author}{Weng, G.}, \bibinfo{author}{Shen, C.}, \bibinfo{author}{Cao, D.}, \bibinfo{author}{Gao, J.}, \bibinfo{author}{Dong, X.}, \bibinfo{author}{He, Q.}, \bibinfo{author}{Yang, B.}, \bibinfo{author}{Li, D.}, \bibinfo{author}{Wu, J.}, \bibinfo{author}{Hou, T.}, \bibinfo{year}{2021}.
\newblock \bibinfo{title}{{PROTAC-DB: an online database of PROTACs}}.
\newblock \bibinfo{journal}{Nucleic acids research} \bibinfo{volume}{49}, \bibinfo{pages}{D1381--D1387}.
\bibitem[{Winter et~al.(2019)Winter, Montanari, Steffen, Briem, No{\'e} and Clevert}]{winter2019efficient}
\bibinfo{author}{Winter, R.}, \bibinfo{author}{Montanari, F.}, \bibinfo{author}{Steffen, A.}, \bibinfo{author}{Briem, H.}, \bibinfo{author}{No{\'e}, F.}, \bibinfo{author}{Clevert, D.A.}, \bibinfo{year}{2019}.
\newblock \bibinfo{title}{Efficient multi-objective molecular optimization in a continuous latent space}.
\newblock \bibinfo{journal}{Chemical Science} \bibinfo{volume}{10}, \bibinfo{pages}{8016--8024}.
\bibitem[{Wu et~al.(2021)Wu, Zhu, Kang, Leung, Lei, Shen, Jiang, Wang, Cao and Hou}]{wu2021we}
\bibinfo{author}{Wu, Z.}, \bibinfo{author}{Zhu, M.}, \bibinfo{author}{Kang, Y.}, \bibinfo{author}{Leung, E.L.H.}, \bibinfo{author}{Lei, T.}, \bibinfo{author}{Shen, C.}, \bibinfo{author}{Jiang, D.}, \bibinfo{author}{Wang, Z.}, \bibinfo{author}{Cao, D.}, \bibinfo{author}{Hou, T.}, \bibinfo{year}{2021}.
\newblock \bibinfo{title}{{Do we need different machine learning algorithms for QSAR modeling? A comprehensive assessment of 16 machine learning algorithms on 14 QSAR data sets}}.
\newblock \bibinfo{journal}{Briefings in Bioinformatics} \bibinfo{volume}{22}, \bibinfo{pages}{bbaa321}.

\end{thebibliography}

\appendix
\section*{Appendices}

\section{Prediction Scores}
\label{appendix:scores}

\begin{figure*}[b]
    \centering
    \begin{subfigure}{0.5\textwidth}
        \centering
        \includegraphics[width=0.99\columnwidth]{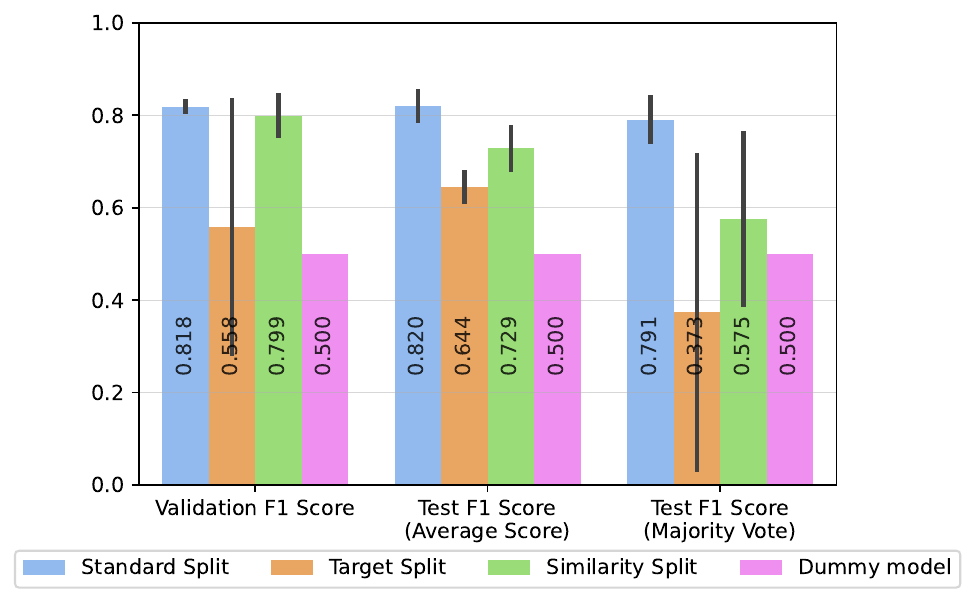}
        \caption{}
        \label{fig:cellsonehot_f1_score}
    \end{subfigure}%
    \begin{subfigure}{0.5\textwidth}
        \centering
        \includegraphics[width=0.99\columnwidth]{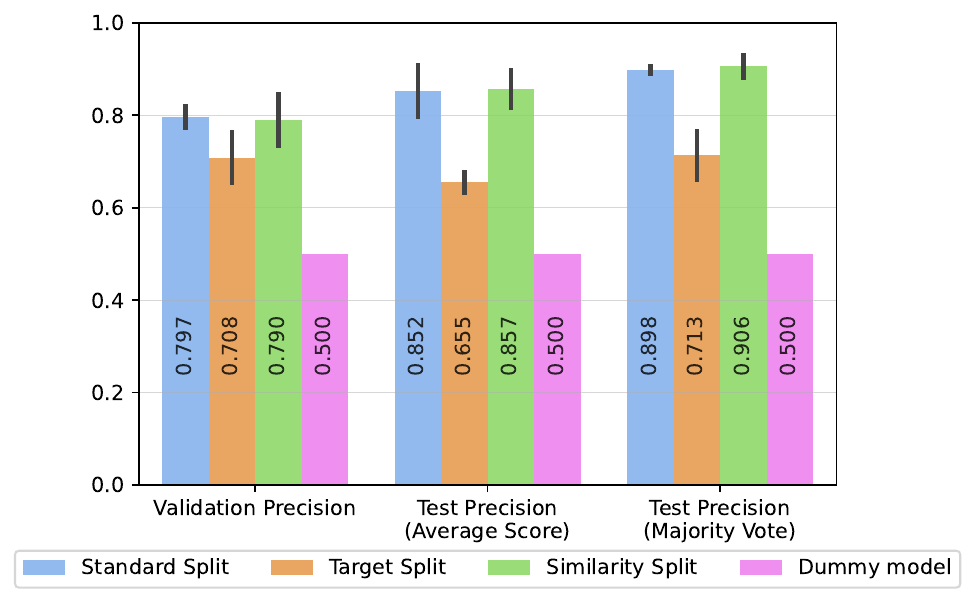}
        \caption{}
        \label{fig:cellsonehot_precision}
    \end{subfigure}\\%
    \begin{subfigure}{0.5\textwidth}
        \centering
        \includegraphics[width=0.99\columnwidth]{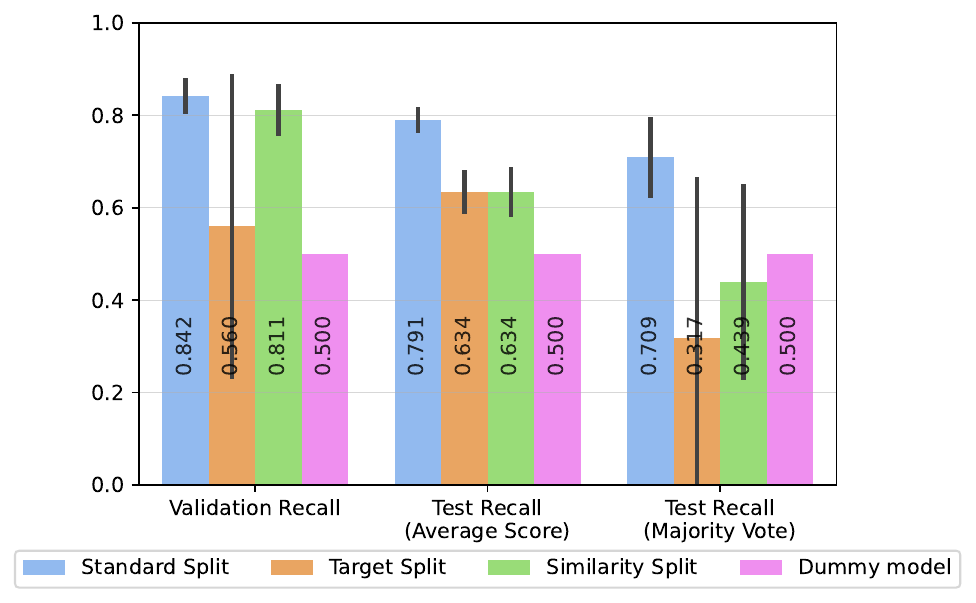}
        \caption{}
        \label{fig:cellsonehot_recall}
    \end{subfigure}%
    \caption{Additional performance metrics for the presented deep learning models: (a) F1 score, (b) precision, and (c) recall.}
    \label{fig:cellsonehot_performance}
\end{figure*}

This section provides a collection of all the validation and test scores computed for the evaluated models on the three proposed studies, always comparing to a dummy model that simply predicts the majority class in the training set.
Figures \ref{fig:cellsonehot_f1_score}, \ref{fig:cellsonehot_precision}, and \ref{fig:cellsonehot_recall} report validation and test F1, precision, and recall scores, respectively, of the evaluated deep learning models in this work. Validation and test accuracy and ROC-AUC scores are reported in Figure \ref{fig:performance} in the main text. 

\section{Cell Line Embeddings From Text Descriptions}
\label{appendix:cell_line_embeddings}
This section details the methods used to extract complex cell line embedding vectors from text descriptions.
A basic approach is to assign a categorical (or one-hot encoded) label to each cell line in the dataset.
While practical, this method ignores any inherent information about the cell lines and their biological similarity.
To address this, we utilized the Cellosaurus database, which provides standardized information about common cell lines used in research \citep{bairoch2018cellosaurus}.
Our approach involves isolating relevant biological information about each cell line into a text description. Features such as omics, genome ancestry, doubling time, and sequence variations, all in text form, are ranked by uniqueness and filtered to form a concise single text description of a given cell line.
We then encode this text into an embedding vector by using a sentence Transformer model \citep{reimers-2019-sentence-bert}.

\begin{figure*}[H!]
    \centering
    \begin{subfigure}{0.46\textwidth}
        \centering
        \includegraphics[width=0.99\columnwidth]{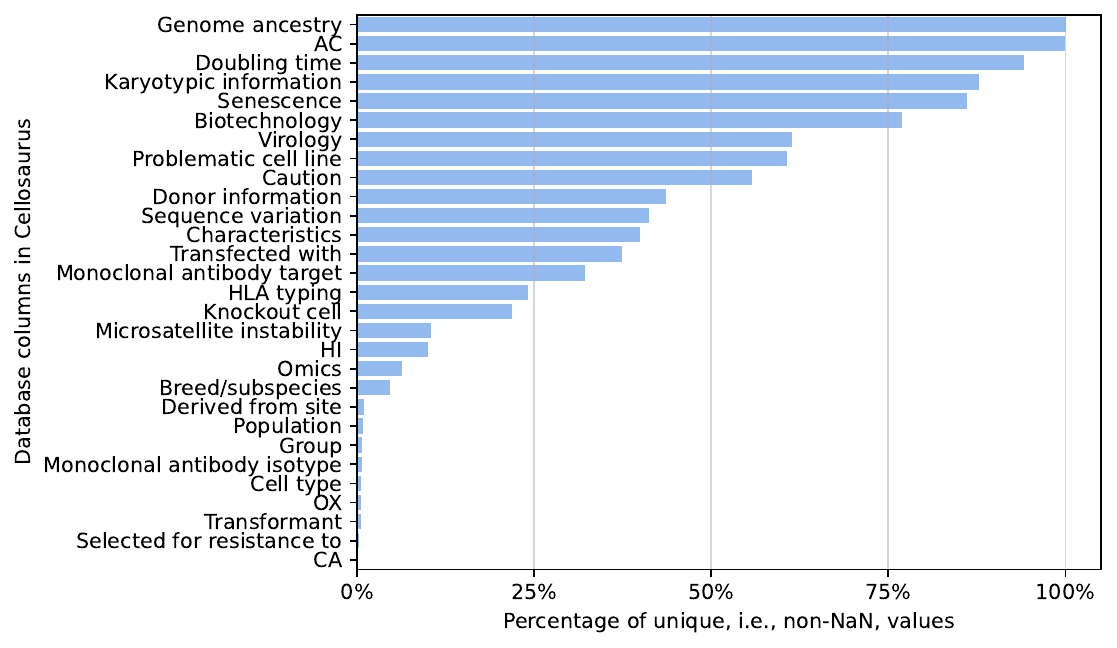}
        \caption{}
        \label{fig:cell_unique_cols}
    \end{subfigure}%
    \begin{subfigure}{0.54\textwidth}
        \centering
        \includegraphics[width=0.99\columnwidth]{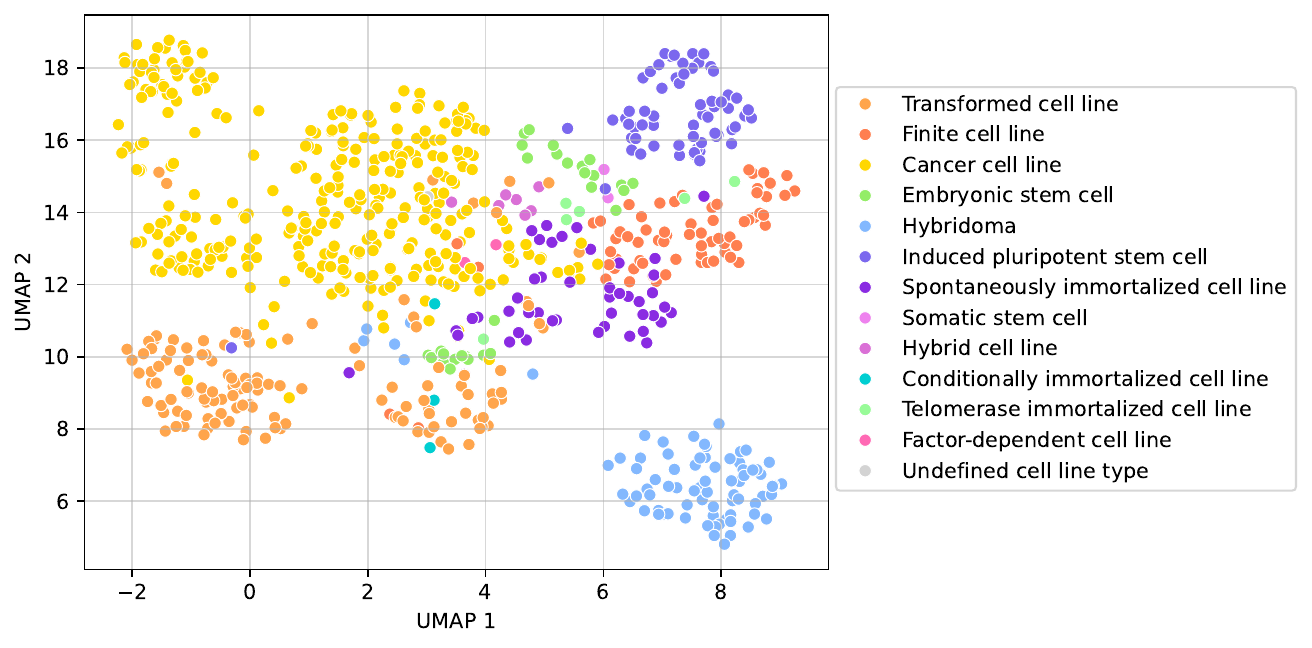}
        \caption{}
        \label{fig:cell_umap}
    \end{subfigure}%
    \caption{(a) Cell line information (database columns) from Cellosaurus, ranked by their percentage of unique entries over the total number of entries in that column. (b) UMAP visualization of the generated cell line embedding vectors, color-coded by cell line categories.}
    \label{fig:cells}
\end{figure*}

A sentence Transformer is designed to generate embedding representations of input sentences such that similar sentences have high cosine similarity.
However, sentence Transformers have a fixed input size, accepting a maximum number of tokens.
To process longer texts, we divide them into chunks of the maximum size, encode each chunk into a vector, and average the vectors into a single representation.
To avoid diluting relevant information during this averaging process, we aim to summarize each cell line's information into concise, yet informative, short text descriptions.

We manually isolated columns containing relevant biological information about cells from the available database columns, such as their category (\eg, ``hybridoma'', ``cancer cell line'', ``transformed cell line''), sex (male or female), and species of origin (\eg, ``mus musculus'', ``homo sapiens'', etc.).
We discarded identification information, such as patents, synonyms, or entry dates.
Additionally, Cellosaurus provides comments in various categories (\eg, ``monoclonal antibody target'', ``sequence variation'', etc.), which we also included.
The list of selected information is shown on the y-axis of Figure \ref{fig:cell_unique_cols}.

Next, we ranked columns and comments based on the fraction of unique entries relative to their total, as illustrated in Figure \ref{fig:cell_unique_cols}.
Our intuition is that comments with a high number of unique entries help identify specific cell lines, making it easier to distinguish cell types.
Following this principle and after reviewing examples, we selected the following information in this order: genome ancestry, karyotypic information, senescence, biotechnology, virology, caution, donor information, sequence variation, characteristics, transfected with, monoclonal antibody target, HLA typing, knockout cell, microsatellite instability, hierarchy (HI), breed/subspecies, derived from site, population, group, monoclonal antibody isotype, cell type, transformant, selected for resistance to, and category (CA).

Finally, for each database entry, we concatenated the strings from the selected information, removed PubMed references, and stripped extra spaces. The average text description length (\ie, number of characters) of the cell lines in our curated dataset was 181.1, below the 384-token input size limit of the selected sentence Transformer model.

\subsection{Cosine Similarity of Cell Line Descriptions}

Table \ref{tab:example_cell_cosine_similarity} presents a cosine similarity matrix for three cell line descriptions generated by following the above methodology. The cosine similarity metric quantifies the similarity between the textual descriptions of different cell lines, with values ranging from 0 to 1, where 1 indicates identical descriptions and 0 indicates no similarity.

For instance, the description of the cell line \textit{UKF-NB-2rDACARB4} is highly similar to that of \textit{UKF-NB-2rDOCE10}, with a cosine similarity of 0.8759. Both of these cell lines are cancer cell lines derived from the same species (\textit{Homo sapiens}) and are part of the resistant cancer cell line (RCCL) collection. They differ primarily in their resistance to different chemotherapeutic agents: dacarbazine for \textit{UKF-NB-2rDACARB4} and docetaxel for \textit{UKF-NB-2rDOCE10}.

In contrast, the description of \textit{FHS036i-sh18961C}, an induced pluripotent stem cell line, has a much lower similarity to the cancer cell lines, with cosine similarities of 0.2832 and 0.3522 to \textit{UKF-NB-2rDACARB4} and \textit{UKF-NB-2rDOCE10}, respectively. This lower similarity is expected given the fundamental differences in cell type, collection origin, and specific biological characteristics.

These examples illustrate how cosine similarity can effectively differentiate between cell lines based on their detailed descriptions, reflecting both broad classifications and specific attributes.

\begin{table*}[ht!]
\centering
\caption{Example of cosine similarity matrix for three cell line descriptions.}
\begin{tabularx}{\textwidth}{>{\raggedright\arraybackslash}Xccc}
\toprule
\textbf{Cell Line Text Description} & \textbf{UKF-NB-2rDACARB4} & \textbf{UKF-NB-2rDOCE10} & \textbf{FHS036i-sh18961C} \\ \midrule
\texttt{\textbf{UKF-NB-2rDACARB4}: CVCL\_RT02, Cancer cell line, NCBI\_TaxID=9606; ! Homo sapiens (Human), Part of: Resistant Cancer Cell Line (RCCL) collection, Selected for resistance to: ChEBI; CHEBI:4305; Dacarbazine (DTIC; (5-(3,3-dimethyl-1-triazeno)imidazole-4-carboxamide)), Derived from site: Metastatic; Bone marrow; UBERON=UBERON\_0002371, NCIt; C3270; Neuroblastoma, ORDO; Orphanet\_635; Neuroblastoma, CVCL\_9902 ! UKF-NB-2} & 1.0000 & 0.8759 & 0.2832 \\ \hline
\texttt{\textbf{UKF-NB-2rDOCE10}: CVCL\_RR83, Cancer cell line, NCBI\_TaxID=9606; ! Homo sapiens (Human), Part of: Resistant Cancer Cell Line (RCCL) collection, Selected for resistance to: ChEBI; CHEBI:4672; Docetaxel anhydrous (Taxotere), Derived from site: Metastatic; Bone marrow; UBERON=UBERON\_0002371, NCIt; C3270; Neuroblastoma, ORDO; Orphanet\_635; Neuroblastoma, CVCL\_9902 ! UKF-NB-2, Cancer cell line} & 0.8759 & 1.0000 & 0.3522 \\ \hline
\texttt{\textbf{FHS036i-sh18961C}: CVCL\_YY67, Induced pluripotent stem cell, NCBI\_TaxID=9606; ! Homo sapiens (Human), Part of: Framingham Heart Study (FHS) collection, Part of: Next Generation Genetic Association studies (Next Gen) program cell lines, Population: Caucasian, Sequence variation: Mutation; HGNC; 3231; CELSR2; Simple; c.*919G; dbSNP=rs12740374; Zygosity=Homozygous; Note=Major haplotype (PubMed=28388431), Omics: Transcriptome analysis by RNAseq, Derived from site: In situ; Peripheral blood; UBERON=UBERON\_0000178, CVCL\_YY66 ! FHS035i-sh18961A} & 0.2832 & 0.3522 & 1.0000 \\ \bottomrule
\end{tabularx}
\label{tab:example_cell_cosine_similarity}
\end{table*}

\subsection{UMAP Visualization of Cell Line Embeddings}

Figure \ref{fig:cell_umap} presents a uniform manifold approximation and projection (UMAP) plot of the cell line embedding vectors. UMAP is a dimensionality reduction technique that helps visualize high-dimensional data by projecting it into a lower-dimensional space, preserving both local and global data structure \citep{mcinnes2020umapuniformmanifoldapproximation}.

The plot showcases the embedding vectors of cell lines, color-coded according to their categories. Each point represents a cell line, and its position reflects the similarity of its embedding vector to others. Similar cell lines cluster together, indicating that the embedding vectors effectively capture meaningful biological relationships. For instance, induced pluripotent stem cells (light purple) and hybridoma cell lines (light blue) form distinct, dense clusters, demonstrating the embeddings' ability to reflect their biological differences. In contrast, some categories, such as spontaneously immortalized cell lines (purple) and cancer cell lines (yellow), exhibit partial overlap, suggesting shared biological features while maintaining enough distinction to form identifiable subclusters. This visual validation underscores the embeddings' capacity to encapsulate and differentiate between various cell line categories, supporting the efficacy of our approach.

\subsection{Evaluation}

\begin{figure*}[H!]
    \centering
    \begin{subfigure}{0.49\textwidth}
        \centering
        \includegraphics[width=0.99\columnwidth]{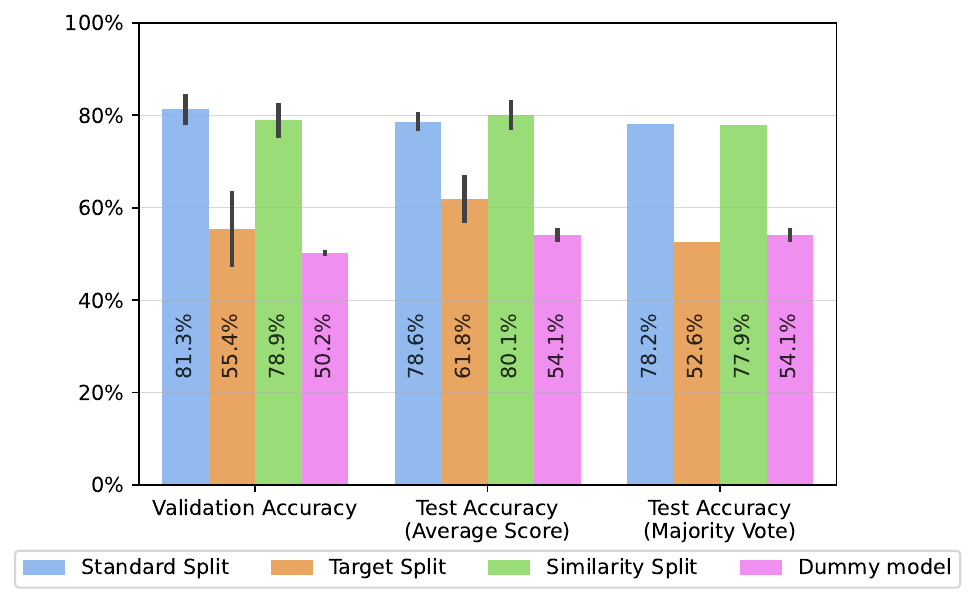}
        \caption{}
        \label{fig:pytorch_accuracy}
    \end{subfigure}%
    \begin{subfigure}{0.49\textwidth}
        \centering
        \includegraphics[width=0.99\columnwidth]{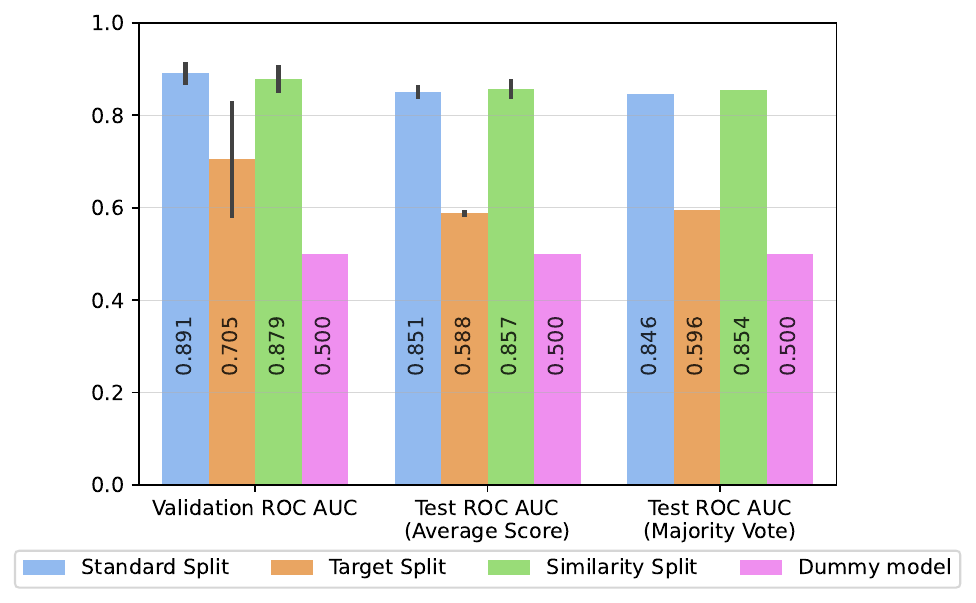}
        \caption{}
        \label{fig:pytorch_roc_auc}
    \end{subfigure}%
    \caption{Performance of the models leveraging cell embeddings from cell line descriptions: (a) accuracy, and (b) ROC-AUC.}
    \label{fig:pytorch_performance}
\end{figure*}

Figure \ref{fig:pytorch_performance} reports the evaluation scores of models leveraging cell line embeddings from text descriptions.
The models were tuned and trained following the same methodology described in Section \ref{sec:hyperparameter_tuning_and_ablation}.
Compared to the baseline model scores on the standard split study in Figure \ref{fig:cellsonehot_performance}, using text-based cell line embeddings resulted in a drop in validation average/test average/test majority vote accuracy of 0.1\%/2.2\%/1.3\%.
Notably, on the similarity split study the model reached validation average/test average/test majority vote accuracy of 78.9\%/80.1\%/77.9\%, compared to 79.6\%/74.9\%/67.5\% for the baseline.
However, the models trained on text-based cell embeddings still struggle to generalize against new targets in the target split study, displaying slightly lower accuracy than the baseline model on this same split (validation average/test average/test majority vote accuracy of 55.4\%/61.8\%/52.6\%).

\section{Datasets Characteristics}

\subsection{PROTAC-DB and PROTAC-Pedia}

Table \ref{tab:dataset-characteristics} provides an overview of the two datasets used in our study: PROTAC-DB and PROTAC-Pedia. PROTAC-DB contains a total of 5,388 entries, whereas PROTAC-Pedia comprises 1,203 entries. The number of unique SMILES in PROTAC-DB is 3,270, compared to 1,178 in PROTAC-Pedia. Unique targets in PROTAC-DB and PROTAC-Pedia are 323 and 79, respectively.
A notable proportion of SMILES entries are shared between the datasets. Specifically, there are 1,222 SMILES entries that are found in both PROTAC-DB and PROTAC-Pedia. These shared SMILES are present in 22.7\% of the total SMILES entries in PROTAC-DB and in 69.2\% of the total SMILES entries in PROTAC-Pedia.
The datasets also feature entries with SMILES that appear only once, \ie, ``single'' SMILES, with 45.5\% (2,451) of PROTAC-DB and 95.8\% (1,153) of PROTAC-Pedia consisting of single SMILES. Single targets are relatively low in both datasets, at 1.4\% (78) for PROTAC-DB and 1.2\% (15) for PROTAC-Pedia, which is unsurprising as multiple PROTACs are generally investigated for a given target.

\begin{table*}[t!]
    \centering
    \caption{Characteristics of PROTAC-DB and PROTAC-Pedia datasets. The term \textit{single} here indicates entries for which the SMILES or target appears only once in the corresponding dataset.}
    \resizebox{\textwidth}{!}{
    \begin{tabular}{lrrrrlrlrl}
    \toprule
    \textbf{Dataset} &  \textbf{Total Entries} &  \textbf{Unique SMILES} &  \textbf{Unique Targets} &  \textbf{Shared SMILES} & \textbf{Shared SMILES \%} &  \textbf{Single SMILES} & \textbf{Single SMILES \%} &  \textbf{Single Targets} & \textbf{Single Targets \%} \\
    \midrule
           PROTAC-DB &             5388 &                    3270 &                      323 &                    1222 &                    22.7\% &                    2451 &                    45.5\% &                       78 &                      1.4\% \\
        PROTAC-Pedia &             1203 &                    1178 &                       79 &                     832 &                    69.2\% &                    1153 &                    95.8\% &                       15 &                      1.2\% \\
    \bottomrule
    \end{tabular}
    }
    \label{tab:dataset-characteristics}
\end{table*}

\begin{table*}[t!]
    \centering
    \caption{Statistics of datasets used in different studies. The term \textit{leaking} indicates the percentage of entries in the training set with either a SMILES or target that also appears in the test set data samples. The \textit{avg Tanimoto distance} refers to the average Tanimoto distance between PROTACs in the test set.}
    \resizebox{\textwidth}{!}{
    \begin{tabular}{rlrrrllllll}
    \toprule
     \textbf{Fold} & \textbf{Study split} &  \textbf{Train size} &  \textbf{Val size} &  \textbf{Test size} & \textbf{Train active \%} & \textbf{Val active \%} & \textbf{Test active \%} & \textbf{Leaking Uniprot \%} & \textbf{Leaking SMILES \%} & \textbf{Avg Tanimoto distance} \\
    \midrule
                              0 &             Standard &                  560 &                140 &                  78 &                   49.6\% &                 50.0\% &                  55.1\% &                      79.1\% &                      8.8\% &                          0.379 \\
             1 &             Standard &                  560 &                140 &                  78 &                   49.6\% &                 50.0\% &                  55.1\% &                      80.0\% &                      8.4\% &                          0.379 \\
             2 &             Standard &                  560 &                140 &                  78 &                   49.6\% &                 50.0\% &                  55.1\% &                      80.2\% &                      9.3\% &                          0.379 \\
             3 &             Standard &                  560 &                140 &                  78 &                   49.8\% &                 49.3\% &                  55.1\% &                      79.3\% &                      8.6\% &                          0.379 \\
             4 &             Standard &                  560 &                140 &                  78 &                   49.8\% &                 49.3\% &                  55.1\% &                      79.3\% &                      7.1\% &                          0.379 \\ \midrule
             0 &               Target &                  594 &                108 &                  76 &                   51.3\% &                 41.7\% &                  53.9\% &                       0.0\% &                      1.0\% &                          0.390 \\
             1 &               Target &                  559 &                143 &                  76 &                   47.2\% &                 60.1\% &                  53.9\% &                       0.0\% &                      0.9\% &                          0.390 \\
             2 &               Target &                  556 &                146 &                  76 &                   46.8\% &                 61.6\% &                  53.9\% &                       0.0\% &                      1.4\% &                          0.390 \\
             3 &               Target &                  592 &                110 &                  76 &                   50.3\% &                 47.3\% &                  53.9\% &                       0.0\% &                      1.2\% &                          0.390 \\
             4 &               Target &                  507 &                195 &                  76 &                   53.8\% &                 39.5\% &                  53.9\% &                       0.0\% &                      1.2\% &                          0.390 \\ \midrule
             0 &           Similarity &                  571 &                110 &                  75 &                   52.7\% &                 49.1\% &                  48.0\% &                      58.7\% &                      0.0\% &                          0.412 \\
             1 &           Similarity &                  546 &                135 &                  75 &                   53.5\% &                 46.7\% &                  48.0\% &                      63.0\% &                      0.0\% &                          0.412 \\
             2 &           Similarity &                  534 &                147 &                  75 &                   49.6\% &                 61.2\% &                  48.0\% &                      59.0\% &                      0.0\% &                          0.412 \\
             3 &           Similarity &                  563 &                118 &                  75 &                   51.5\% &                 55.1\% &                  48.0\% &                      59.7\% &                      0.0\% &                          0.412 \\
             4 &           Similarity &                  510 &                171 &                  75 &                   53.3\% &                 48.5\% &                  48.0\% &                      60.0\% &                      0.0\% &                          0.412 \\
    \bottomrule
    \end{tabular}
    }
    \label{tab:studies_datasets_stats}
\end{table*}

\subsection{Cross-Validation Folds and Test Sets}

Table \ref{tab:studies_datasets_stats} presents detailed statistics for the datasets used in the three studies proposed in our evaluation strategy.

For the standard split, each fold consists of 560 training entries, 140 validation entries, and 78 test entries. The proportion of active data samples in these splits is consistent and balanced across folds, with the training and validation sets containing around 50\% active samples, and the test set 55.1\%. Notably, a significant percentage of entries have leaking Uniprot identifiers (around 80\%) and a smaller proportion have leaking SMILES (around 8\%). The average Tanimoto distance between PROTACs in the test set is 0.379, indicating moderate structural similarity.

The target split aims to evaluate model generalization to unseen POIs. The training set sizes vary between 507 and 594, with the validation set sizes ranging from 108 to 195, and the test set consistently containing 76 entries. 
Because of stratified folds, the active data proportions in the training, validation, and test sets vary more widely than in the standard split.
In fact, there are no leaking Uniprot identifiers in this split, and the proportion of leaking SMILES is below 1.4\%. The average Tanimoto distance between PROTACs in the test set is slightly higher at 0.390.

For the similarity split, designed to test generalization to new PROTACs, the training set sizes range from 510 to 571, validation sets from 110 to 171, and the test set again consistently contains 75 entries. The active sample proportion in the training sets average around 52.2\%, with the validation set showing slightly more variation. The leaking Uniprot identifiers are around 60.1\%, and there are no leaking SMILES, by construction. The average Tanimoto distance between PROTACs in the test set is the highest among the splits at 0.412, reflecting the structural novelty of the test PROTACs in this specific study.

\section{XGBoost Performance}
\label{appendix:xgboost}

\begin{figure*}[t!]
    \centering
    \begin{subfigure}{0.5\textwidth}
        \centering
        \includegraphics[width=0.99\columnwidth]{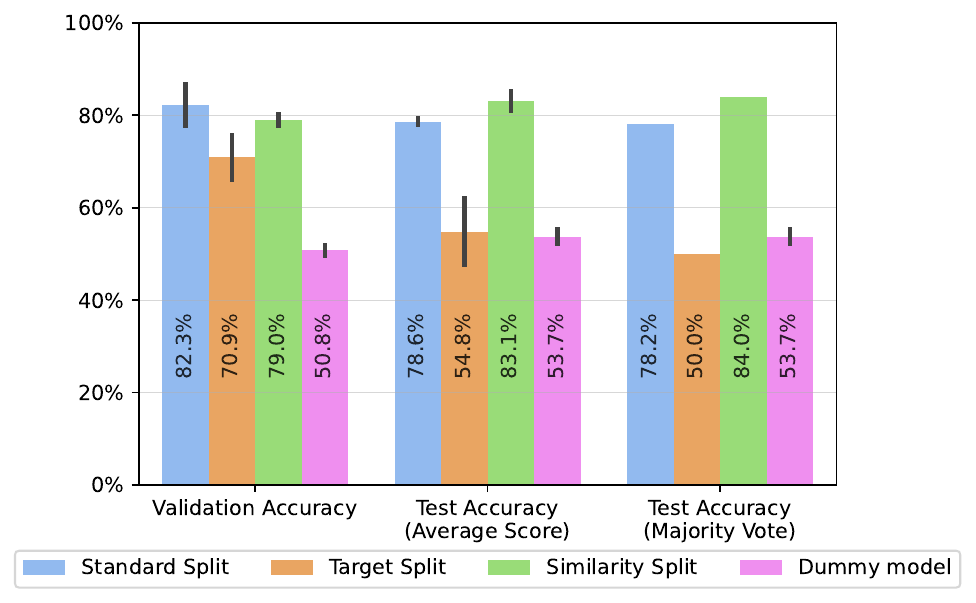}
        \caption{}
        \label{fig:xgboost_accuracy}
    \end{subfigure}%
    \begin{subfigure}{0.5\textwidth}
        \centering
        \includegraphics[width=0.99\columnwidth]{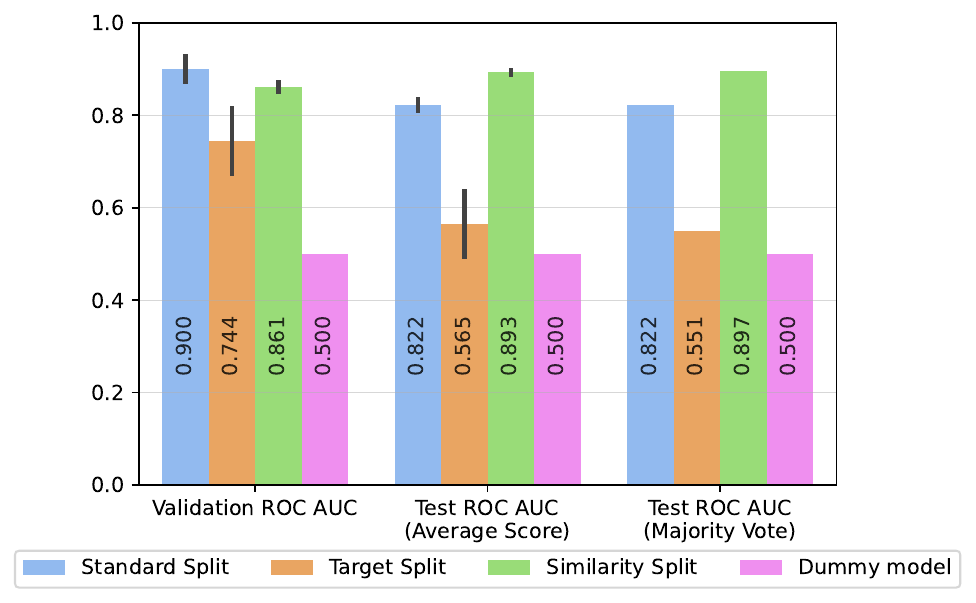}
        \caption{}
        \label{fig:xgboost_roc_auc}
    \end{subfigure}\\%
    \begin{subfigure}{0.5\textwidth}
        \centering
        \includegraphics[width=0.99\columnwidth]{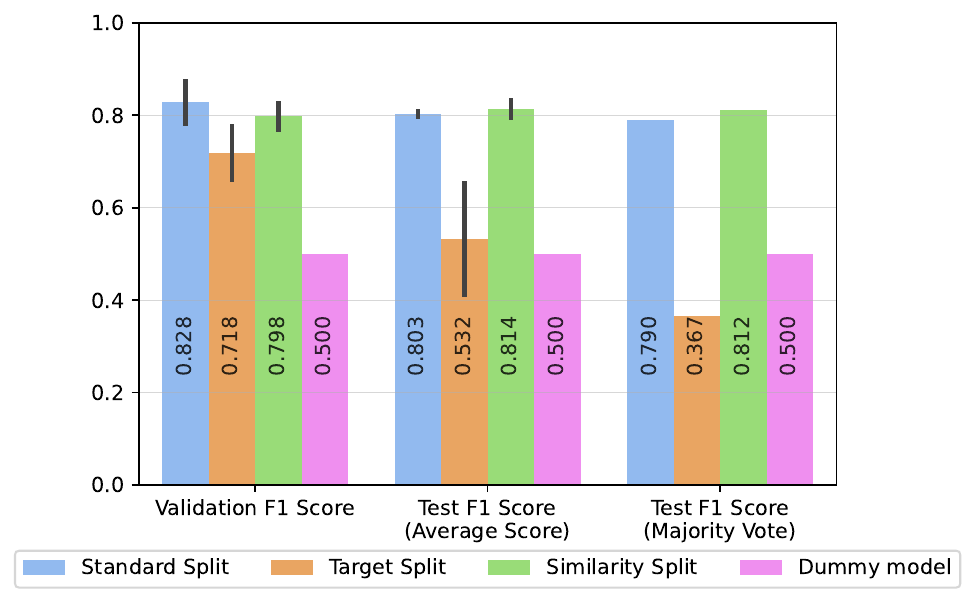}
        \caption{}
        \label{fig:xgboost_f1_score}
    \end{subfigure}%
    \begin{subfigure}{0.5\textwidth}
        \centering
        \includegraphics[width=0.99\columnwidth]{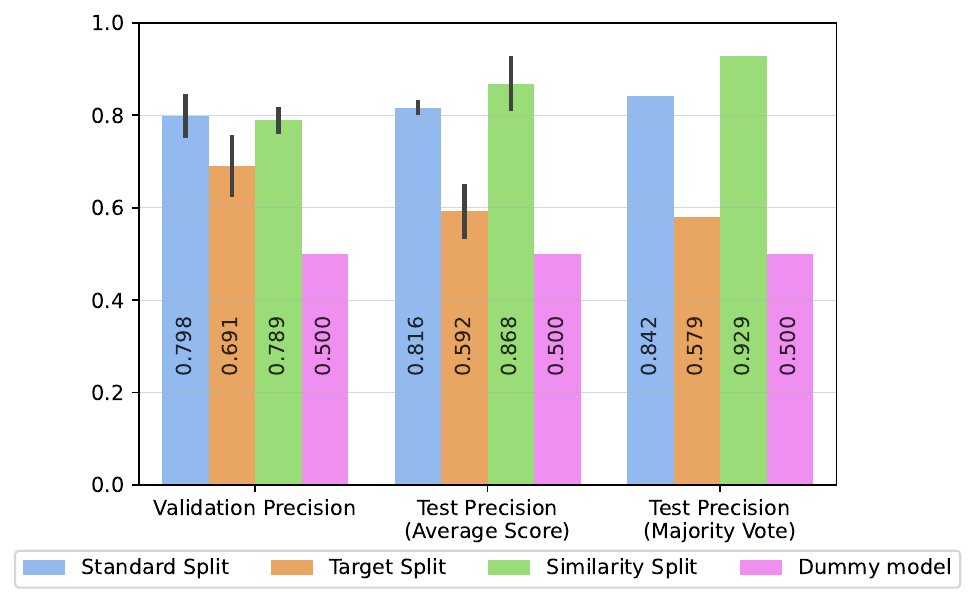}
        \caption{}
        \label{fig:xgboost_precision}
    \end{subfigure}\\%
    \begin{subfigure}{0.5\textwidth}
        \centering
        \includegraphics[width=0.99\columnwidth]{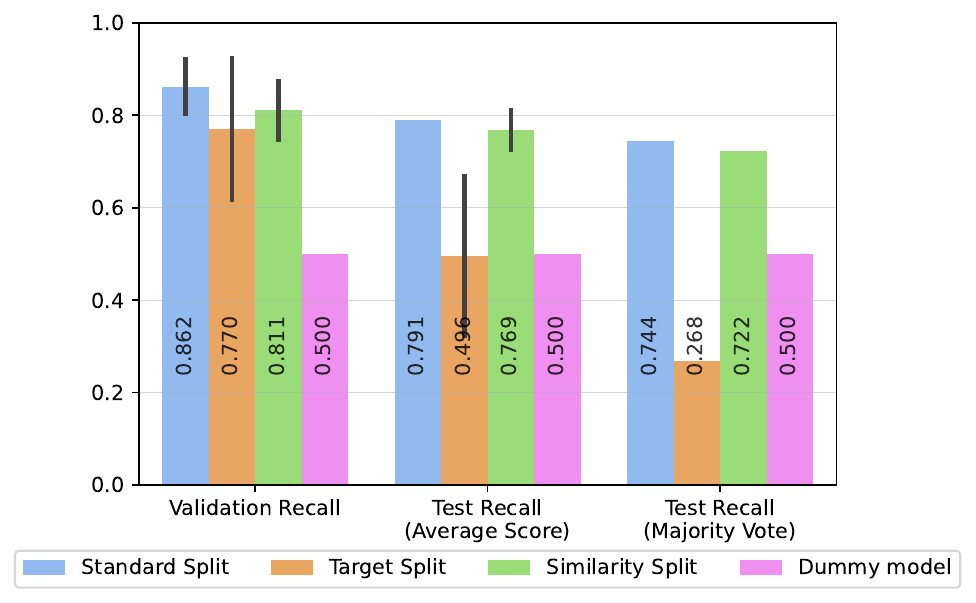}
        \caption{}
        \label{fig:xgboost_recall}
    \end{subfigure}%
    \caption{Performance metrics for the XGBoost models: (a) accuracy, (b) ROC-AUC, (c) F1 score, (d) precision, and (e) recall.}
    \label{fig:xgboost_performance}
\end{figure*}

Given the experimental setup and evaluation strategy described in Section \ref{sec:evaluation_strategy}, we first trained different XGBoost models in a CV setting via Optuna.
The selected hyperparameters tuned in Optuna are reported in Table \ref{tab:xgboost_hyperparameters}.
We then trained, with the best hyperparameters found, three models and evaluated them on the held-out test sets.
As with the deep learning models, we evaluated the XGBoost models both individually by computing their average performance, and together via majority voting.
Figure \ref{fig:xgboost_performance} compares the performance metrics for the trained XGBoost models on the different studies.

The comparison of test performances between the trained XGBoost models and the proposed deep learning models highlights some key differences across the various studies. In the standard random split, deep learning models achieve slightly higher test accuracies (up to 80.08\%) compared to XGBoost (up to 78.6\%). For the target split, deep learning models outperform XGBoost with test accuracies ranging from 55.3\% to 62.3\%, while XGBoost's performance is significantly lower and more variable, ranging from 50\% to 54.8\%.
Curiously, in the similarity split, deep learning models show slightly worse performance with accuracies reaching up to 74.9\% compared to XGBoost’s 83.1\%.

Regarding ROC-AUC scores, in the standard split, both models perform robustly, but deep learning models have slightly higher scores (up to 0.865) compared to XGBoost’s 0.822. In the target split, deep learning models have a clear advantage with ROC-AUC scores up to 0.616, while XGBoost’s scores hover around 0.55, indicating poor performance. In the similarity split, XGBoost models demonstrate better performance with ROC-AUC scores up to 0.897, compared to deep learning’s 0.850. Overall, deep learning models generally show superior or comparable test performance, especially in the target and standard splits.

\begin{table*}[h!]
\centering
\caption{XGBoost hyperparameters tuned in Optuna.}
\resizebox{\textwidth}{!}{
\begin{tabular}{llll}
\toprule
\textbf{Parameter} & \textbf{Type} & \textbf{Range} & \textbf{Scale} \\ \midrule
Step size shrinkage (\texttt{eta}) & float & 1e-4 to 1e-1 & log \\ 
Maximum depth of a tree (\texttt{max\_depth}) & int & 3 to 10 &  \\ 
Minimum sum of instance weight (hessian) needed in a child (\texttt{min\_child\_weight}) & float & 1e-3 to 10.0 & log \\ 
Minimum loss reduction required to make a further partition (\texttt{gamma}) & float & 1e-4 to 1e-1 & log \\ 
Subsample ratio of the training instances (\texttt{subsample}) & float & 0.5 to 1.0 &  \\ 
Subsample ratio of columns when constructing each tree (\texttt{colsample\_bytree}) & float & 0.5 to 1.0 &  \\ \bottomrule
\end{tabular}
}
\label{tab:xgboost_hyperparameters}
\end{table*}

\section{Amino Acid Counts as Protein Embeddings}
\label{appendix:protein_embeddings_as_aa_counts}

\begin{figure*}[H!]
    \centering
    \begin{subfigure}{0.49\textwidth}
        \centering
        \includegraphics[width=0.99\columnwidth]{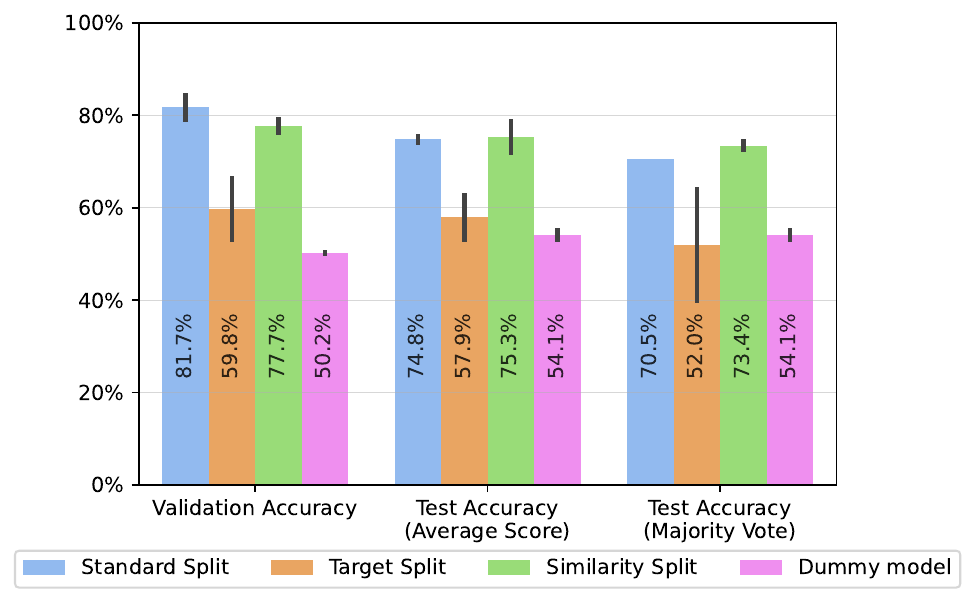}
        \caption{}
        \label{fig:aminoacidcnt_accuracy}
    \end{subfigure}%
    \begin{subfigure}{0.49\textwidth}
        \centering
        \includegraphics[width=0.99\columnwidth]{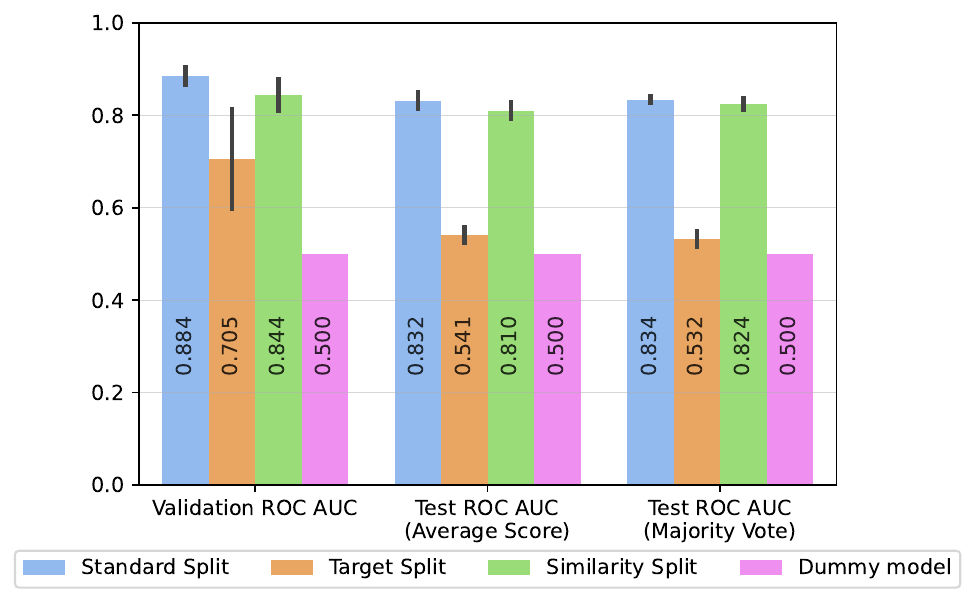}
        \caption{}
        \label{fig:aminoacidcnt_roc_auc}
    \end{subfigure}%
    \caption{Performance of the models leveraging amino acid counts as protein embeddings: (a) accuracy, and (b) ROC-AUC.}
    \label{fig:aminoacidcnt_performance}
\end{figure*}

We conducted an experiment to investigate whether our model utilizes latent information from protein structure embeddings or merely uses them as ``barcodes'' to differentiate between data samples.
We encoded both the POI and E3 ligase amino acid sequences with a 1-gram count-vectorizer from scikit-learn, \ie, we counted the characters in the sequence string.
We then proceeded to tune and train the models following the same methodology detailed in Section \ref{sec:hyperparameter_tuning_and_ablation}.
Figure \ref{fig:aminoacidcnt_performance} shows the obtained performance scores.
When encoding protein sequences as amino acid counts, we can see validation average/test average/test majority vote accuracy differences of +0.3\%/-6\%/-9\%, compared to the baseline model scores showed in Figure \ref{fig:cellsonehot_performance}.
In particular, the models completely fail to generalize against new targets, reaching a top test accuracy of 57.9\%.
Overall, the embeddings from pretrained Transformers used by our models, although not perfect, appear to help the models learn more meaningful latent representations of the biological context of PROTACs.

\end{document}